\documentclass{vgtc}                          % final (conference style)

\graphicspath{{figures/}{pictures/}{images/}{./}} % where to search for the images

\usepackage{times}                    % we use Times as the main font
\usepackage{tabu}                      % only used for the table example
\usepackage{booktabs}                  % only used for the table example
\usepackage{lipsum}                    % used to generate placeholder text
\usepackage{mwe}                       % used to generate placeholder figures
\usepackage{amsmath,amsfonts}
\usepackage{algorithmic}
\usepackage{algorithm}
\usepackage{array}
\usepackage[caption=false,font=normalsize,labelfont=sf,textfont=sf]{subfig}
\usepackage{textcomp}
\usepackage{stfloats}
\usepackage{url}
\usepackage{verbatim}
\usepackage{graphicx}
\usepackage{cite}
\usepackage{bbding}
\usepackage{amssymb}
\usepackage{soul}
\usepackage{makecell}
\usepackage{colortbl}

\usepackage{mathptmx}                  % use matching math font

\onlineid{1068}
\vgtccategory{Research}
\vgtcinsertpkg

\usepackage{soul,color,xcolor}      % 用于设置字体颜色
\definecolor{myColor}{rgb}{0.8039,0,0}   % 显示修订时用这行
\makeatletter
\newcommand*{\revise}{\@ifnextchar\bgroup{\revise@}{\color{myColor}}}
\newcommand*{\revise@}[1]{{\textcolor{myColor}{#1}}}
\makeatother

\title{ArchGPT: Understanding the World's Architectures \\with Large Multimodal Models}

\author{%
Yuze Wang\textsuperscript{1}\quad
Luo Yang\textsuperscript{1}\quad
Junyi Wang\textsuperscript{2}\quad
Yue Qi\textsuperscript{1}
}

\affiliation{%
\scriptsize
\textsuperscript{1} State Key Laboratory of Virtual Reality Technology and Systems, School of Computer Science and Engineering, Beihang University\\
\textsuperscript{2} School of Computer Science and Technology, Shandong University
}

\teaser{
  \centering
  \includegraphics[width=\linewidth]{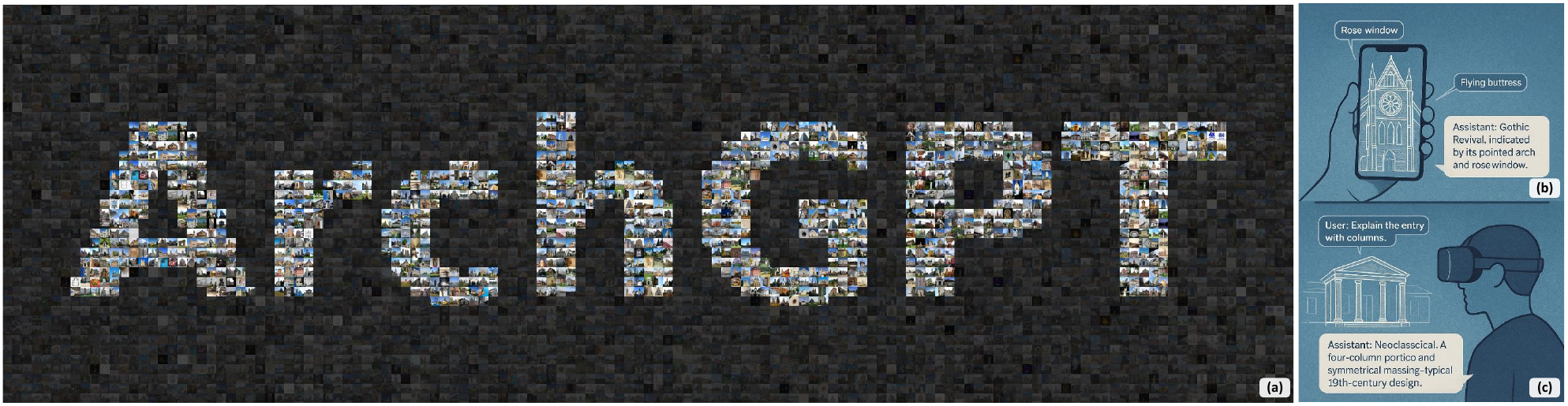}
  \caption{  \textbf{Teaser.} We present ArchGPT, a domain-adapted multimodal model for architecture-specific visual analysis. Panel (a) shows a photomosaic from the proposed Arch-300K dataset that powers ArchGPT. ArchGPT can seamlessly integrates into XR applications—(b) AR: interactive guided tours with augmented overlays and natural-language queries; (c) VR: ask-as-you-explore virtual walkthroughs with real-time architectural Q\&A. }
  \label{fig:teaser}
}

\abstract{
Architecture embodies aesthetic, cultural, and historical values, standing as a tangible testament to human civilization. Researchers have long leveraged virtual reality (VR), mixed reality (MR), and augmented reality (AR) to enable immersive exploration and interpretation of architecture, enhancing accessibility, public understanding, and creative workflows around architecture in education, heritage preservation, and professional design practice. However, existing VR/MR/AR systems are often developed case-by-case, relying on hard-coded annotations and task-specific interactions that do not scale across diverse built environments. In this work, we present ArchGPT, a multimodal architectural visual question answering (VQA) model, together with a scalable data-construction pipeline for curating high-quality, architecture-specific VQA annotations. This pipeline yields Arch-300K, a domain-specialized dataset of approximately 315{,}000 image-question-answer triplets. Arch-300K is built via a multi-stage process: first, we curate architectural scenes from Wikimedia Commons and filter unconstrained tourist photo collections using a novel coarse-to-fine strategy that integrates 3D reconstruction and semantic segmentation to select occlusion-free, structurally consistent architectural images. To mitigate noise and inconsistency in raw textual metadata, we propose an LLM-guided text verification and knowledge-distillation pipeline to generate reliable, architecture-specific question-answer pairs. Using these curated images and refined metadata, we further synthesize formal analysis annotations-including detailed descriptions and aspect-guided conversations-to provide richer semantic variety while remaining faithful to the data.
We perform supervised fine-tuning of an open-source multimodal backbone ,ShareGPT4V-7B, on Arch-300K, yielding ArchGPT. Through quantitative and qualitative evaluations, robustness analyses, ablation studies, and VR-based interactive demonstrations, we show that ArchGPT delivers robust multimodal reasoning and analysis across diverse architectural aspects, enabling scalable and interactive architecture-aware VR experiences. We will release the dataset and model weights to support future research.
}


\begin{document}
\newcommand{\DicArch}{D_{A}}
\newcommand{\DicNonarch}{D_{N}}

\newcommand{\Image}{I}
\newcommand{\Conf}{C}
\newcommand{\MaskC}{M^{c}}
\newcommand{\MaskF}{M^{f}}
\newcommand{\ratioC}{r^{c}}
\newcommand{\ratioF}{r^{f}}
\newcommand{\PortionCoarse}{\tau_c}
\newcommand{\PortionFine}{\tau_f}

\newcommand{\ThreeDimModel}{P}
\newcommand{\AlphaCoarse}{\alpha_c}

\newcommand{\FuncVGG}{F_{VGG}}

\newcommand{\SceneName}{\texttt{\{scene\_name\}}}
\newcommand{\WikiRawData}{\texttt{\{wiki\_raw\_data\}}}
\newcommand{\SceneTax}{\texttt{\{scene\_taxonomy\}}}
\newcommand{\SceneYear}{\texttt{\{scene\_year\}}}
\newcommand{\SceneLocation}{\texttt{\{scene\_location\}}}
\newcommand{\SceneRefinedDesc}{\texttt{\{scene\_refined\_description\}} }
\newcommand{\SceneFormalName}{\texttt{\{scene\_formal\_name\}}}
\firstsection{Introduction}

\maketitle

Architecture encodes aesthetic, cultural, and historical knowledge in built form, serving as a vital record of human civilization.
In recent years, virtual reality (VR), mixed reality (MR), and augmented reality (AR) technologies have increasingly supported the study and presentation of architectural heritage.
Prior work explores architectural reconstruction~\cite{vrarch_recon_1,agarwal2011building,NeuralRecon-W} and virtual exploration~\cite{nerf-w,we-gs,wang2025look,kulhanek2024wildgaussians}, enabling users to navigate historical buildings from arbitrary viewpoints with photorealistic detail.
Other efforts~\cite{vrarch_st_1,vrarch_st_2,wu2021towers,wang2025taking} integrate historical and stylistic narratives into immersive environments to enhance cultural engagement.
These applications improve accessibility, public understanding, and creative workflows around architecture.
Nevertheless, most existing systems remain case-by-case solutions with hardcoded annotations and task-specific interaction flows, and they cannot interpret visual input or support open-ended interaction at scale.

Recent advances in large multimodal models (LMMs), such as LLaVA~\cite{li2024llava}, GPT-4V~\cite{openai2023gpt4v}, and Gemini~\cite{team2023gemini}, enable open-ended reasoning over visual inputs.
Could these advances inspire us to build a domain-adapted model for architecture, providing a foundation for diverse applications?
While these models perform strongly on general visual question answering, they struggle with architecture-specific understanding.
A key issue is their tendency toward \emph{large language model (LLM)-biased visual hallucination}~\cite{huang2025survey_hallucination}, a phenomenon in which the model misidentifies a scene and then analyzes the image based on the wrong recognition, relying more on language priors than on the pixels.
This \emph{recognize-then-analyze} behavior undermines reliability in domains that require grounded visual analysis.
The core limitation lies not in model architecture, but in the lack of curated, structured architectural data during pretraining.
Most LMMs are trained on internet-scale image-text pairs dominated by natural scenes and object-centric captions, with little exposure to annotated architectural imagery that captures stylistic features, material usage, symbolic motifs, or cultural context.
This data-domain misalignment yields brittle performance even on basic questions such as "What is the architectural style?" or "What architectural elements are visible in the facade?"

Constructing an architecture-specific visual question answering (VQA) dataset is non-trivial: fine-grained annotation demands time, labor, and expert knowledge.
To address this challenge, we design a multi-stage dataset construction pipeline that leverages LLM knowledge priors with multimodal web metadata to generate high-quality, domain-specific VQA annotations.
Specifically, we collect high-quality architectural images and metadata textual descriptions from Wikimedia Commons~\cite{wikimediacommons}.
To ensure data quality, we propose a coarse-to-fine filtering strategy using VGGT~\cite{vggt} and SAM~\cite{sam} to select high-quality and occlusion-free samples from unconstrained image collections.
We then prompt LLMs to produce textual analyses centered on visual characteristics, guided by contextual cues (building name, location, construction date, and historical background) obtained from Wikimedia textual data.
To enrich diversity, the pipeline is instructed to structure outputs across multiple perspectives, including architectural style, elements, context, symbolism, and materials.

This process covers about 9{,}000 scenes and yields roughly 315{,}000 high-quality VQA pairs and descriptive entries, consolidated into a new dataset named \textbf{Arch-300K}.
Building on this dataset, we present \textbf{ArchGPT}, a domain-adapted multimodal model tailored for architecture-specific VQA and fine-grained visual interpretation.
ArchGPT follows a LLaVA-style architecture and is supervised fine-tuned from ShareGPT4V-7B.
We compare against strong open-source LMM baselines and demonstrate robust, adaptable architectural understanding. Finally, we showcase three applications for ArchGPT: an interactive architectural conversational assistant; AR-enabled immersive architectural interpretation; and ask-as-your-explore virtual walkthroughs.

In summary, our contributions are threefold:

$\bullet$ \textbf{Methodology.} We propose a multi-stage dataset-construction pipeline that leverages LLM knowledge priors with multimodal web metadata to generate high-quality, domain-specific VQA annotations.

$\bullet$ \textbf{Dataset and Model.} We introduce Arch-300K dataset and ArchGPT, a domain-adapted LMM fine-tuned on this dataset. We will release the dataset and model weights to support future research.

$\bullet$ \textbf{Insights and applications.}  We provide empirical insights on domain adaptation for LMMs and demonstrate practical use cases in VR/AR environments.
\section{Related Work}
\label{sec:2_related}
\begin{figure*}[!htb]
  \centering
\includegraphics[width=2.0\columnwidth]{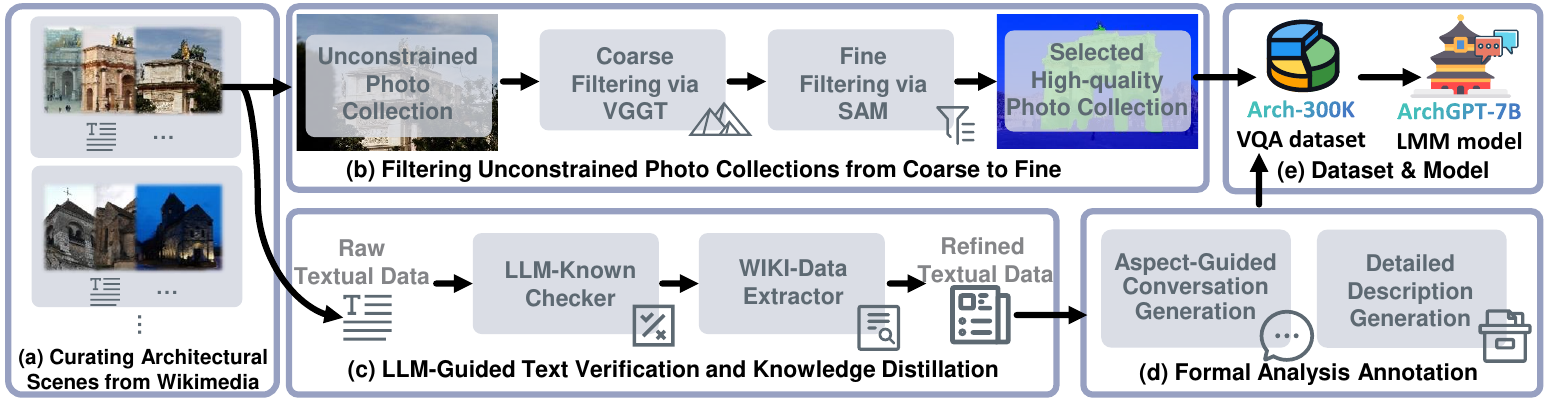}
\caption{The overall pipeline of our method. 
(a) We begin by curating architectural scenes from Wikimedia Commons. 
(b) For each scene, unconstrained photo collections uploaded by tourists are processed through a coarse-to-fine filtering strategy that combines 3D reconstruction and segmentation to select high-quality, occlusion-free images. 
(c) To handle noisy raw textual metadata collected online, we introduce LLM-guided text verification and knowledge distillation, which extract refined architectural descriptions, locations, and scene formal names. 
(d) Using curated images and refined metadata, we generate formal analysis annotations consisting of detailed descriptions and aspect-guided conversations. 
(e) Together, these steps yield the Arch-300K dataset. Building upon Arch-300K, we further fine-tune ShareGPT4V-7B to obtain ArchGPT.}
  \label{fig:method_main}
\end{figure*}
\subsection{Virtual Reality for Architecture}
In architecture, VR, AR, and MR have become versatile media supporting a wide spectrum of activities~\cite{archvr_survey_1,archvr_survey_2,archvr_survey_3,archvr_survey_4}. 
Heritage-oriented works employ photogrammetry~\cite{arch_phtogrammertry_1,arch_phtogrammertry_2}, laser scanning~\cite{arch_laser_1,arch_laser_2}, and computer-aided design (CAD)/ building information modeling (BIM)~\cite{arch_hbim_1,arch_hbim_2,arch_hbim_3,arch_hbim_4} to reconstruct monuments and enable virtual exploration of fragile or inaccessible sites. 
More recent methods leverage neural rendering techniques such as neural radiance fields (NeRFs)~\cite{nerf_vanilla} and 3D Gaussian Splatting (3DGS)~\cite{3dgs} to achieve photorealistic walkthroughs from unconstrained images~\cite{nerf-w,we-gs,wang2025look,kulhanek2024wildgaussians}, expanding the fidelity of architectural VR experiences. 
Beyond geometry and appearance modeling, recent studies have also explored semantic field reconstruction, where models are augmented with closed-set semantics~\cite{wu2021towers}, segment anything (SAM) features~\cite{sam}, and vision-language features (e.g., CLIP~\cite{radford2021learning}) to recognize and decompose architectural components, enabling both visual realism and semantic interpretability in VR environments.
Complementing reconstruction efforts, prior work embeds historical and stylistic narratives within immersive experiences-via AR overlays and VR storytelling-to enhance cultural engagement and education~\cite{arch_vr_story_1,arch_vr_story_2,arch_vr_story_3,arch_vr_story_4}.
Beyond these efforts, VR/AR/MR platforms have also been adopted in architectural practice, supporting immersive modeling, collaborative design feedback, and retrofitting\cite{arch_vr_app_1,arch_vr_app_2,arch_vr_app_3}. 

Together, these works demonstrate the value of VR as a medium for both preservation and creative practice in architecture. 
However, most existing systems remain case-specific, requiring hand-crafted scripts or annotations tailored to individual sites or projects, which fundamentally limits scalability. 
In contrast, our work pursues a generalizable framework for open-ended architectural analysis, moving beyond case-by-case methods.

\subsection{Large Language Models}
The past several years have witnessed remarkable progress in LLMs, driven by the transformer architecture~\cite{transformer} and large-scale pretraining frameworks exemplified by BERT~\cite{bert}, GPT-3~\cite{gpt3}, and GPT-4~\cite{gpt4}. A key milestone for the research community was the release of the LLaMA family~\cite{llama,llama2}, which provided open-access checkpoints at multiple parameter scales. Its efficiency and accessibility have made LLaMA the de facto backbone for many instruction-tuned and multimodal derivatives, accelerating innovation across the LLM ecosystem.

Within the architectural domain, recent LLM applications have focused on building codes and regulations, compliance checking, and documentation. For example, PB-ACC~\cite{yang2024prompt} automates the transformation of building-code information to simplify compliance checks; ITcon-Safety~\cite{tran2024leveraging} extracts and structures construction safety regulations; LegalRuleML-LLM~\cite{fuchs2024using} translates building regulations into computable LegalRuleML; and DAVE~\cite{fernandes2024gpt} introduces a GPT-powered assistant that enables real-time multimodal interaction with BIM data via text or voice.

These efforts demonstrate the utility of LLMs for rule interpretation and compliance workflows. In contrast, our work targets the aesthetic and visual dimensions of architecture, supporting open-ended analysis of style, form, and contextual cues.

\subsection{Large Multimodal Models}
The rapid growth of LLMs has catalyzed research on vision-language interaction, with early work such as CLIP~\cite{radford2021learning} aligning images and text via contrastive learning. More recent methods (e.g., MiniGPT-4~\cite{zhu2023minigpt}, LLaVA~\cite{li2024llava}) extend LLMs with visual adapters to support image-text dialogue through improved pretraining and supervised fine-tuning.
General-purpose LMMs, including ShareGPT4V~\cite{chen2024sharegpt4v}, Qwen-VL~\cite{llm_5}, and InternVL~\cite{zhu2025internvl3}, integrate visual features using learnable projectors or query embeddings. They rely on multimodal pretraining with instruction tuning to handle broad vision-language tasks across domains.
Domain-specialized LMMs show the value of targeted data and objectives: LLaVA-Med~\cite{llavamed} adapts to biomedical imagery and dialogue; GalleryGPT~\cite{bin2024gallerygpt} produces paragraph-level analyses for paintings; and ArtiMuse~\cite{cao2025artimuse} advances aesthetic scoring with expert-style interpretability.

Despite this progress, the architectural domain lacks a public, large-scale VQA dataset covering parts, styles, materials, and typologies. We address this gap with constructing Arch-300K dataset and an architecture-adapted LMM, ArchGPT, that performs open-ended, evidence-grounded architectural analysis. To our knowledge, ArchGPT is the first multimodal model explicitly specialized for architecture, moving beyond general-purpose assistants toward domain-specific architectural reasoning.

\section{The Proposed Method}
\label{sec:method}

Constructing large-scale, high-quality VQA datasets for the architectural domain is challenging due to the need for expert knowledge and the scarcity of structured supervision. An overview of our pipeline is shown in \cref{fig:method_main}.
First, we curate image collections and metadata from Wikimedia Commons~\cite{wikimediacommons} and identify architectural scenes (\cref{sec:3_dataset_curation}). We then filter high-quality images via a coarse-to-fine process that combines 3D reconstruction with semantic segmentation (\cref{sec:3_data_collection_scene_and_image}).
Second, we perform LLM-guided text verification and knowledge distillation: an LLM-Known checker assesses scene familiarity of LLM, and a Wiki-data extractor normalizes names and extracts refined metadata to guide VQA annotation (\cref{sec:3_data_collection_llm_guided_text_verification_and_knowledge_distillation}).
Third, guided by these priors, we synthesize two types of VQA annotations, detailed descriptions and aspect-guided conversations, and construct the Arch-300K dataset (\cref{sec:3_formal_analysis_annotation}).
Finally, we perform supervised fine-tuning of an open-source multimodal backbone, ShareGPT4V-7B, on Arch-300K to obtain the ArchGPT, a domain-adapted multimodal model for architecture-specific visual analysis (\cref{sec:3_model}).

\subsection{Curating Architectural Scenes from Wikimedia Commons}
\label{sec:3_dataset_curation}

\paragraph{\textbf{Treating Wikimedia Commons Categories as Scenes.}}
We collect scenes from Wikimedia Commons \cite{wikimediacommons}, following the MegaScenes \cite{megascenes} of treating each category on the Wikimedia as a distinct scene. MegaScenes contains approximately 450,000 scenes across various categories, including outdoor, indoor, and natural landscapes.
Building upon MegaScenes, we selectively filter architectural scenes to construct our dataset.
\paragraph{\textbf{Preliminary Filtering of Architectural Scenes.}}
We begin by filtering Wikimedia Commons scenes to identify architectural content. To achieve this, we utilize the scene name and taxonomy information to determine whether it belongs to architecture. We define two dictionaries: $\DicArch$, containing architectural keywords (e.g., "building," "arch," "chapel"), and $\DicNonarch$, containing non-architectural keywords (e.g., "nature," "activity," "organization").
For each scene, we check its name and taxonomy.
If it contains terms from $\DicArch$, we classify it as an architectural scene; if it contains terms from $\DicNonarch$, we classify it as non-architectural.
If neither $\DicArch$ nor $\DicNonarch$ terms are found, we use LLM to assess the scene by querying: "Do you think $\SceneName$, categorized under $\SceneTax$, is an architectural structure? Please respond with Yes or No."
This allows the LLM to provide a final judgment on whether the scene is architectural.
For this task, we use the Gemini 2.5 Pro model \cite{team2023gemini}.

At this stage, approximately 40\% of the scenes are classified as architectural. 
Each architectural scene contains a set of unconstrained images taken by tourists, along with scene-related textual metadata such as the scene name, scene taxonomy, sub-taxonomy, and raw textual descriptions available on Wikimedia Commons.

\subsection{Filtering Unconstrained Photo Collections from Coarse to Fine}
\label{sec:3_data_collection_scene_and_image}
\begin{figure}[t]
  \centering
  \includegraphics[width=\linewidth]{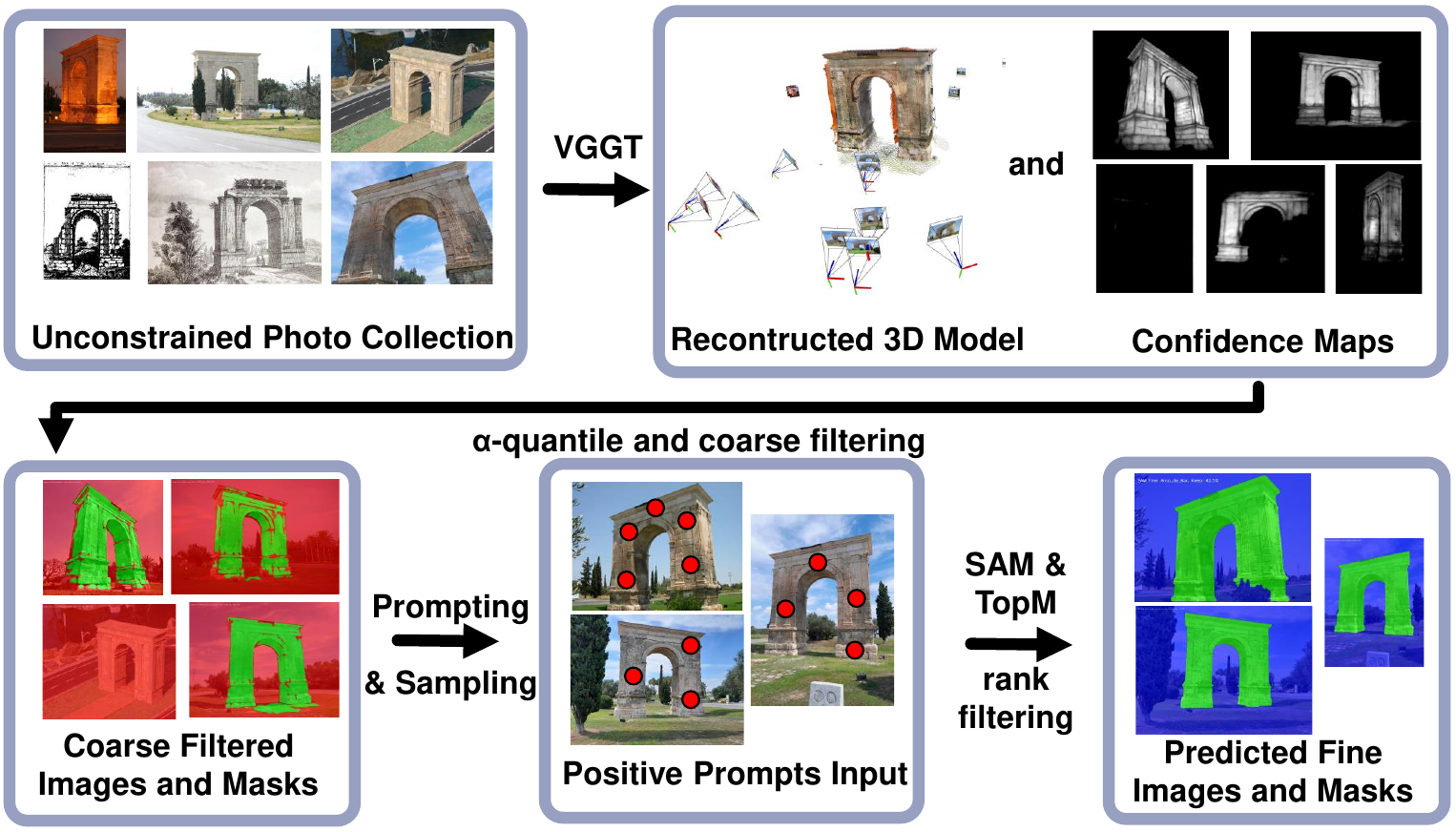}
  \caption{Overview of the coarse-to-fine unconstrained photo collection filtering pipeline. The process combines 3D reconstruction confidence and semantic segmentation to select high-quality, occlusion-free architectural images.}
  \label{fig:method_filtering_pipeline}
\end{figure}

Internet-sourced architectural scenes often include unconstrained photo collections captured from diverse viewpoints, with different cameras, at different times, and under varying weather conditions. 
These collections frequently contain transient occluders such as pedestrians or vehicles, reducing their suitability for architectural analysis.
To address these challenges, we propose a coarse-to-fine image filtering pipeline that selects high-quality, occlusion-free images by combining 3D reconstruction with semantic segmentation, as shown in \cref{fig:method_filtering_pipeline}. 
Our key assumption is that architecture-focused images will exhibit higher 3D reconstruction confidence, and will yield more consistent and complete segmentations.
\paragraph{\textbf{Coarse Filtering via VGGT.}}
Compared with optimization-based pipelines (e.g., COLMAP~\cite{colmap}), feed-forward 3D reconstruction methods run at much lower latency, making reconstruction-driven image filtering feasible at scale across large numbers of scenes.
Given $N$ unconstrained images $\{\Image_1, \Image_2, \ldots, \Image_N\}$ in a scene, we first employ the feed-forward reconstruction model VGGT~\cite{vggt} to generate a 3D model $\ThreeDimModel$ and per-image confidence maps $\{\Conf_1, \Conf_2, \ldots, \Conf_N\}$:
\begin{equation}
\Conf_1, \ldots, \Conf_N, \ThreeDimModel = \FuncVGG(\Image_1, \ldots, \Image_N).
\end{equation}
We additionally apply sky segmentation using ONNX~\cite{onnx}, and ignore sky regions when computing confidence statistics, as they provide little structural information. We then compute the $\alpha$-quantile (set as $\alpha=0.8$) over all pixel-wise confidence values across the image collection to obtain a global confidence threshold $\AlphaCoarse$.
Each confidence map $\Conf_i$ is thresholded to produce a coarse binary mask $\MaskC_i$:
\[
\MaskC_i(x,y) =
\begin{cases}
1, & \text{if } \Conf_i(x,y) \ge \AlphaCoarse, \\
0, & \text{otherwise.}
\end{cases}
\]
We calculate the proportion of pixels marked as 1 in each $\MaskC_i$:
\begin{equation}
\ratioC_i = \frac{1}{H \times W} \sum_{x=1}^{H} \sum_{y=1}^{W} \MaskC_i(x,y),
\end{equation}
where $H \times W$ is the image resolution. 
Then, we discard images whose ratio $\ratioC_i$ falls below a threshold $\PortionCoarse$. 
This stage filters out images with low reconstruction reliability or severe occlusion.
\paragraph{\textbf{Fine Filtering via SAM.}}

Since VGGT-based confidence maps can be noisy, we further refine the selected image set with the help of SAM~\cite{sam}. 
For each image $\Image_i$ that passes coarse filtering, we randomly sample $P$ prompt points from regions where $\MaskC_i = 1$. ($P=10$ in our experiments.)
These points are used to prompt SAM to generate a refined architectural mask $\MaskF_i$.
We then compute the valid-pixel ratio for each $\MaskF_i$:
\begin{equation}
\ratioF_i = \frac{1}{H \times W} \sum_{x=1}^{H} \sum_{y=1}^{W} \MaskF_i(x,y).
\end{equation}
The final set of $K$ high-quality images is selected by sorting the images by $\ratioF_i$ and retaining the top $K$:
\begin{equation}
\{\Image_1', \ldots, \Image_K'\} = \operatorname{TopK}\left(\{\ratioF_i\}_{i=1}^{N}, K\right),
\end{equation}
where $\operatorname{TopK}$ ranks images by segmentation coverage and selects the top $K$. We set $K=8$ in our experiments.

This coarse-to-fine procedure enables the selection of high-quality, occlusion-free architectural images from large-scale, unconstrained photo collections. 
During this process, approximately 55\% of the architectural scenes are entirely discarded, as no qualifying images remain after filtering. 
For instance, a scene labeled as an art museum may contain only visitor-uploaded photos of artworks rather than the architecture itself, resulting in poor geometric consistency and complete exclusion by the filtering pipeline.
\begin{figure}[h]
  \centering
  \includegraphics[width=0.8\linewidth]{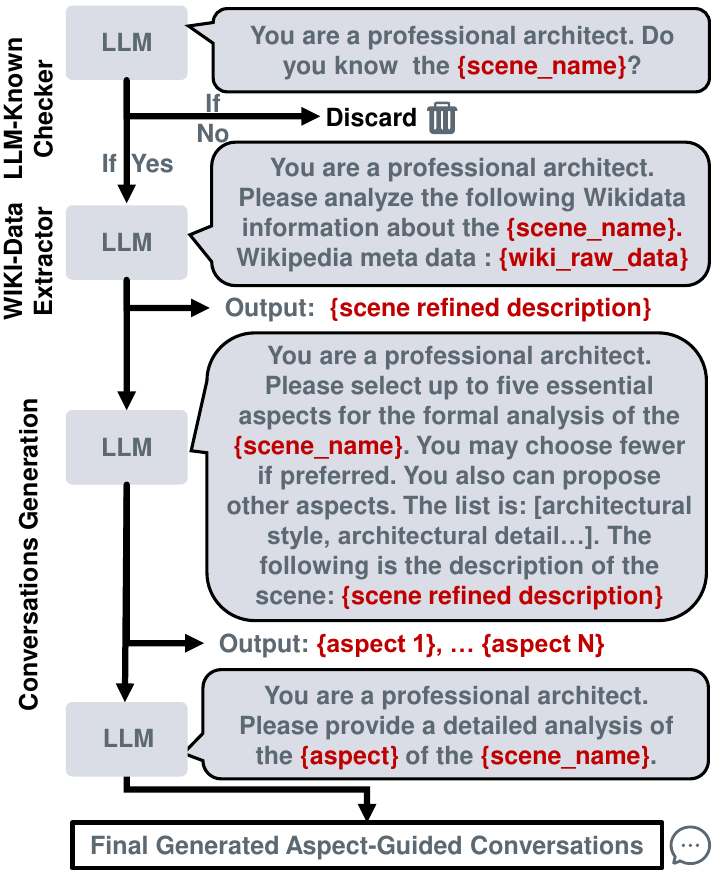}
  \caption{A simplified flowchart of the aspect-guided conversation generation process.}
  \label{fig:method_text}
\end{figure}
\subsection{LLM-Guided Text Verification and Knowledge Distillation}
\label{sec:3_data_collection_llm_guided_text_verification_and_knowledge_distillation}
The core idea in processing textual metadata is to leverage the LLM's prior knowledge: if the model already knows the architecture, then prompting it with curated architectural metadata can elicit rich, structured descriptions-even without the corresponding visual input. 
However, this process faces two main challenges. 
First, the raw textual data extracted from Wikimedia is often noisy and unstructured, requiring additional cleaning and refinement. 
Second, LLMs do not have exhaustive knowledge of all architectural scenes; when prompted with unfamiliar architectures, they may generate inaccurate or misleading content, which degrades annotation quality.
To address these issues, we introduce two components: an LLM-known checker, which verifies whether a scene falls within the LLM's knowledge scope, and a Wiki-data extractor, which filters and distills architecture-relevant information from noisy textual metadata. 
% Together, these components ensure that the final textual descriptions are both reliable and domain-relevant.
In our experiments, we adopt Gemini 2.5 Pro for this step, as it provides more consistent responses and broader architectural knowledge coverage than other publicly available LLMs.
\paragraph{\textbf{LLM-Known Checker.}}
Before extracting architectural knowledge, we first verify whether the scene falls within the LLM's knowledge scope.
Given a scene with $\SceneName$, we issue a binary prompt:
\begin{quote}
"You are a professional architect. Do you know the $\SceneName$? Respond with only `YES' or `NO'."
\end{quote}
If the model responds "NO", we discard the scene and skip annotation; only scenes marked "YES" are retained for subsequent processing.
\paragraph{\textbf{Wiki-Data Extractor.}}
For scenes identified as "known" by the LLM, we combine the model's internal knowledge with supplementary information from Wikimedia Commons-including the scene name $\SceneName$ and associated textual metadata $\WikiRawData$-to extract three key fields:
(i) a refined architectural description $\SceneRefinedDesc$ that captures stylistic and structural characteristics,
(ii) the geographical location $\SceneLocation$ and the construction year $\SceneYear$,
and (iii) the formal architecture name $\SceneFormalName$.
Because Wikimedia scene names are often informal or embellished (e.g., "Arc\_de\_Triomf\_at\_night", "remote\_view\_of\_the\_Arc\_de\_Triomf"), we standardize them into canonical architecture names (e.g., "Arc de Triomf").
All extractions are strictly grounded in the provided textual metadata; if specific information is not present, the output is set to "Unknown."
For more details, please refer to the supplementary materials.

This process can be viewed as retrieval-augmented generation (RAG), where $\WikiRawData$ acts as retrieved context that guides the LLM's reasoning toward structured annotation.
This structured extraction process ensures accurate, consistent, and standardized textual metadata, which in turn improves the quality of VQA annotation and contributes to the overall reliability of our dataset.

\subsection{Formal Analysis Annotation}  
\label{sec:3_formal_analysis_annotation}

\begin{figure*}[t]
  \centering
\includegraphics[width=2.0\columnwidth]{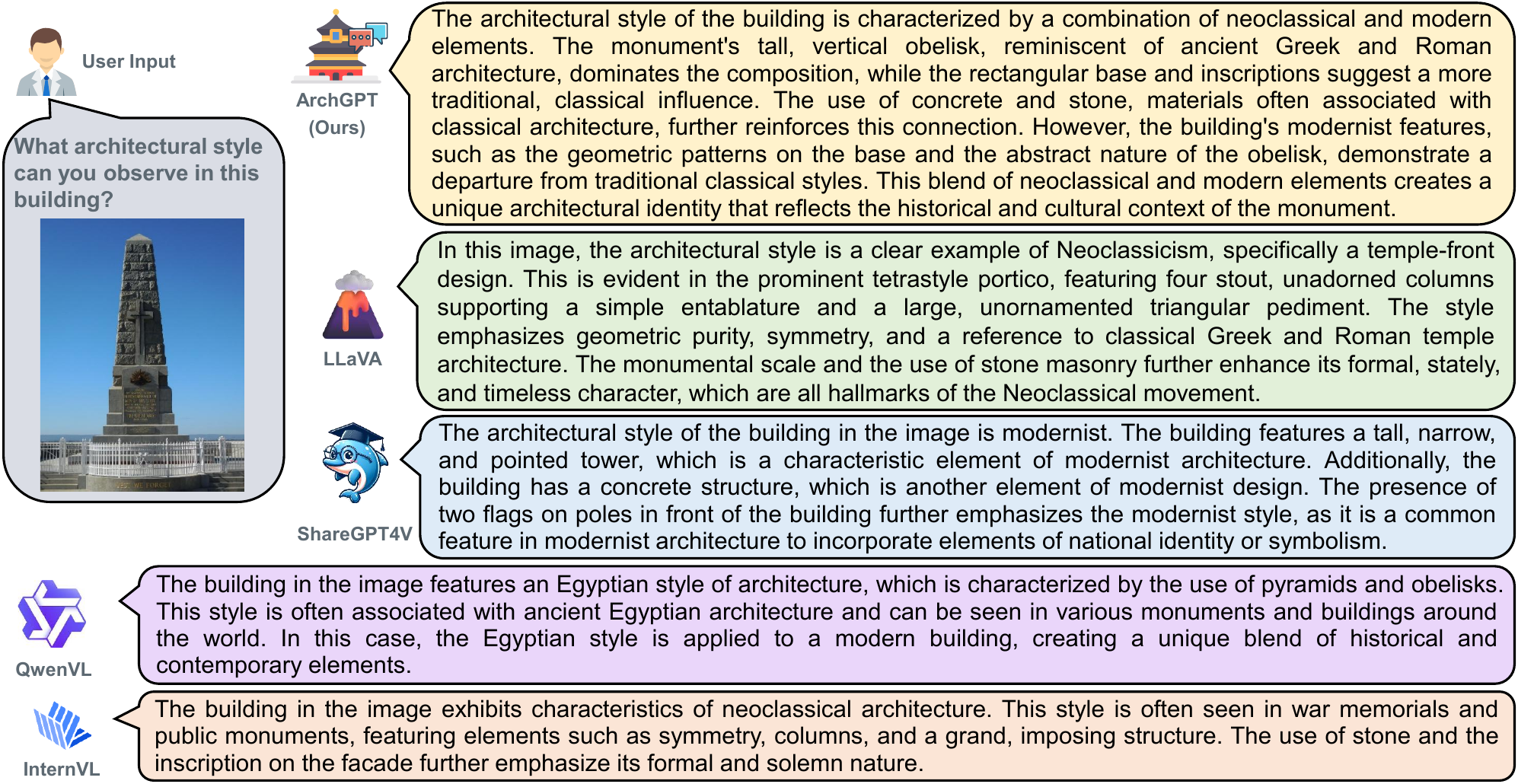}
\caption{Qualitative results about architectural style recognition. ArchGPT provides more accurate, visually grounded, and terminology-precise style attributions than baseline LMMs, with fewer hallucinations.}
  \label{fig:exp_qualitative_results_1}
\end{figure*}

To enable fine-grained architectural reasoning and support diverse vision-language tasks, we construct two types of instruction-following annotations for each architectural scene:
\textbf{Detailed Descriptions}, a paragraph-level narrative summarizing the architecture's overall visual characteristics;
\textbf{Aspect-Guided Conversations}, a set of question-answer pairs focused on specific architectural aspects.
These two formats are designed to complement each other: the former provides a holistic visual overview of the scene, while the latter encourages targeted exploration of architectural dimensions such as style, symbolism, and materials.
\paragraph{\textbf{Detailed Description Generation.}}
For each scene, we construct a structured instruction prompt that guides the LLM to generate visually grounded, expert-style annotations.  
The prompt includes key metadata such as the scene name, location, construction year, and a refined architectural description distilled from textual metadata:
\begin{quote}\small
\texttt{You are a professional architect. Please describe the architecture \{scene\_formal\_name\} in detail. Please compose a coherent paragraph, approximately \{detailed\_words\} words. Start your response with 'In this image, I can see...' or 'This image shows...' Only focus on the VISUAL CHARACTERISTICS. DO NOT mention the specific architecture name. Supplementary material: The architecture's name:\{scene\_formal\_name\}, location: \{scene\_location\},construct year:\{scene\_year\}, other description:\{scene\_refined\_description\}}
\end{quote}

Note that this is a simplified version; the complete prompts are provided in the supplementary materials.
We randomly sample one prompt variant from a curated pool and instruct the LLM to compose a coherent paragraph.  
This scene-level description is then associated with all images within the corresponding scene.
\paragraph{\textbf{Aspect-Guided Conversation Generation.}}
In addition to global descriptions, we also generate question-answer pairs that focus on specific architectural aspects. We provide the LLM with a predefined set of analysis dimensions, including architectural style, architectural elements, architectural details, architectural context, architectural innovation, architectural symbolism, and architectural materials. The model is encouraged to select up to five relevant aspects from this list, but it is also free to incorporate additional dimensions based on its own knowledge of architecture.
For each chosen aspect, the LLM is instructed to (i) formulate a concise, targeted question, and (ii) generate a visually grounded answer. 

The detailed descriptions and aspect-guided conversations offer multi-granularity supervision for downstream architectural understanding and vision-language tasks. 
For more detail, please refer to the supplementary materials. Through our proposed pipeline, we construct a large-scale, high-quality dataset tailored to VQA in the architectural domain.  
After a final round of manual screening and refinement, we construct the dataset comprising 315{,}247 VQA pairs, with 263{,}806 samples for training and 51{,}441 for testing. 
We name this dataset Arch-300K.

\subsection{The Proposed ArchGPT}
\label{sec:3_model}
\paragraph{\textbf{Architecture.}} ArchGPT does not introduce a new multimodal architecture; instead, it focuses on adapting existing LLM for architectural analysis.  
We adopt ShareGPT4V-7B~\cite{chen2024sharegpt4v} as the backbone, following the LLaVA framework~\cite{li2024llava}.  
The overall system consists of three components:
\textbf{Vision Encoder:}  
We use the ViT-L/14 vision transformer from CLIP-Large~\cite{radford2021learning} to encode input images resized to $336 \times 336$.  
The image is divided into $14 \times 14$ patches, producing 576 visual tokens that capture both global structure and local architectural detail.
\textbf{Projection Layer:}  
A lightweight two-layer multilayer perceptron (MLP) projects the visual tokens into the language model's embedding space, enabling effective cross-modal alignment for architecture-specific language reasoning.
\textbf{Language Model:}  
We adopt the 7B variant of Vicuna-v1.5~\cite{chiang2023vicuna}, based on the LLaMA2 decoder-only architecture~\cite{touvron2023llama}.  
We integrate LoRA~\cite{hu2022lora} adapters to adapt the model to the formal style and domain-specific semantics of architectural discourse, while keeping the backbone weights frozen.  
This preserves the general descriptive capability of the pretrained model while enabling efficient domain adaptation.
% This design supports fine-grained architectural reasoning without modifying the backbone architecture.
\paragraph{\textbf{Fine-Tuning.}} We perform supervised fine-tuning of ArchGPT using the constructed Arch-300K dataset.  
To adapt the language model to domain-specific discourse while retaining general-language capability, we attach LoRA adapters (rank $r{=}128$, $\alpha{=}256$) to the ShareGPT4V-7B backbone and freeze the base LLM, preserving its strong descriptive proficiency.
Training is conducted for two epochs on a single GPU with 40 GB memory, using a batch size of 16 and a learning rate of 2e-5. To improve training efficiency, we adopt a multi-annotation fusion strategy inspired by Molmo~\cite{deitke2025molmo}:  
for each image, all associated annotations (descriptions or QA pairs) are concatenated into a single sequence with appropriate attention masks.
\begin{figure*}[t]
  \centering
\includegraphics[width=2.0\columnwidth]{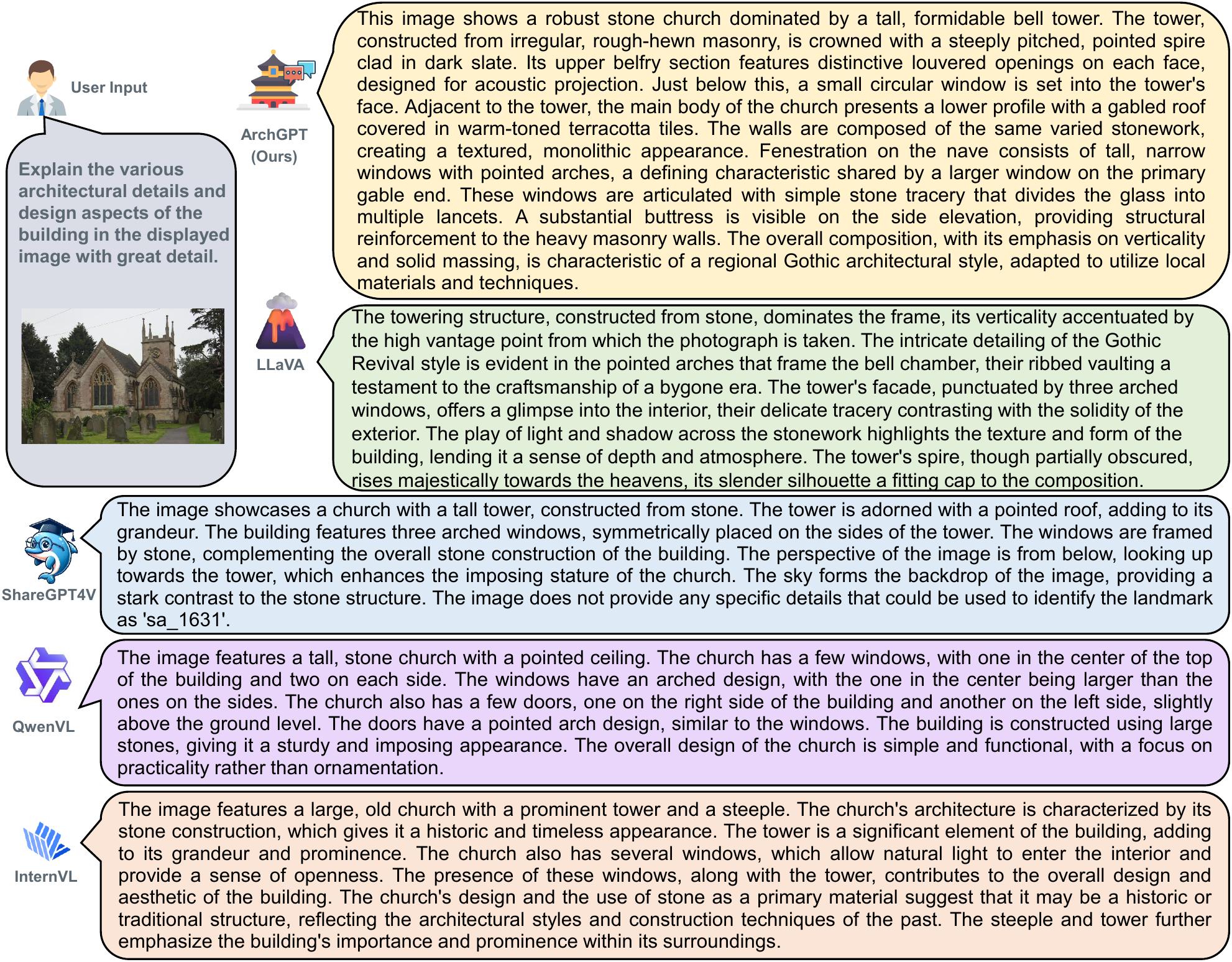}
\caption{Qualitative results about architectural detailed analysis. ArchGPT yields finer-grained, element- and material-aware descriptions with stronger visual grounding and reduced hallucination compared to baselines.}
  \label{fig:exp_qualitative_results_2}
\end{figure*}
\begin{table*}[t]
    \caption{Quantitative experimental results on the proposed Arch-300K dataset, with the first, second, and third values highlighted in red, orange, and yellow, respectively. Our method demonstrates superior overall performance compared to state-of-the-art approaches. $S_{judgeLM}^*$ indicates the score evaluated without architectural emphasis in the prompt, while $S_{judgeLM}$ indicates the vanilla JudgeLM~\cite{zhu2023judgelm}.}
    \label{tab:quantitative_compare_main}
    \centering
    % \resizebox{2.0\columnwidth}{!}{%
    \begin{tabular}{lccccccc}
        \cmidrule[\heavyrulewidth]{1-8}
        Model & $S_{judgeLM}\uparrow$ & $S_{judgeLM}^*\uparrow$ & GLEU$\uparrow$ & METEOR$\uparrow$ & ROUGE-1$\uparrow$ & ROUGE-2$\uparrow$ & ROUGE-L$\uparrow$  \\
        \midrule
        % --- keep these two JudgeLM columns EXACTLY as user provided ---
        LLaVA-1.5-7B~\cite{li2024llava} &
        \cellcolor{orange!25}7.472 & \cellcolor{orange!25}6.779 &
        \cellcolor{orange!25}10.30 & \cellcolor{yellow!25}24.00 & \cellcolor{yellow!25}33.88 & \cellcolor{yellow!25}6.52 & \cellcolor{orange!25}19.96 \\

        ShareGPT4V-7B~\cite{chen2024sharegpt4v} &
        6.348 & 5.624 &
        9.51 & 22.06 & 33.27 & 6.34 & 19.73 \\

        Qwen-VL-Chat-7B~\cite{llm_5} &
        6.182 & 5.490 &
        8.87 & 20.35 & 31.77 & 5.98 & 18.82 \\

        InternVL3-8B~\cite{zhu2025internvl3} &
        \cellcolor{yellow!25}7.150 & \cellcolor{yellow!25}6.554 &
        \cellcolor{yellow!25}9.65 & \cellcolor{orange!25}24.26 & \cellcolor{orange!25}34.44 & \cellcolor{orange!25}7.04 & \cellcolor{yellow!25}19.83 \\
        \midrule
        Ours  &
        \cellcolor{red!25}7.713 & \cellcolor{red!25}7.107 &
        \cellcolor{red!25}14.43 & \cellcolor{red!25}30.45 & \cellcolor{red!25}40.42 & \cellcolor{red!25}12.72 & \cellcolor{red!25}24.91 \\
        \bottomrule
    \end{tabular}
    % }
\end{table*}

This avoids redundant image encoding across annotations and reduces training time by more than half, with negligible sequence overhead. These strategies enable efficient and scalable adaptation of large vision-language models for the architectural domain.

\begin{table*}[!htbp]
\centering
\caption{Aspect-wise JudgeLM results. For each prompt and aspect, JudgeLM evaluates all model outputs and selects the single best response. 
Each cell reports the number of wins (and the corresponding proportion within that aspect), with the highest, second-highest, and third-highest values highlighted in red, orange, and yellow, respectively.
}

\label{tab:judgelm_aspects}
% \resizebox{\linewidth}{!}{
\begin{tabular}{lccccc}
\toprule
Aspect & ArchGPT(Ours) & InternVL3-8B~\cite{zhu2025internvl3} & Qwen-VL-Chat~\cite{llm_5} & LLaVA-LoRA~\cite{llava_benchmark} & ShareGPT4V-7B~\cite{chen2024sharegpt4v} \\
\midrule
Creativity          & \cellcolor{red!25}\textbf{4470 (16.2\%)} & 2743 (9.9\%) & 64 (0.2\%) & \cellcolor{orange!25}4450 (16.1\%) & \cellcolor{yellow!25}4115 (14.9\%) \\
Level of detail     & \cellcolor{red!25}\textbf{6178 (22.4\%)} & 4766 (17.2\%) & 71 (0.3\%) & \cellcolor{yellow!25}5403 (19.6\%) & \cellcolor{orange!25}5597 (20.3\%) \\
Logical consistency & \cellcolor{red!25}\textbf{5990 (21.7\%)} & \cellcolor{yellow!25}4703 (17.0\%) & 72 (0.3\%) & 4575 (16.6\%) & \cellcolor{orange!25}5137 (18.6\%) \\
Domain expertise    & \cellcolor{red!25}\textbf{5006 (18.1\%)} & \cellcolor{yellow!25}3509 (12.7\%) & 56 (0.2\%) & 3333 (12.1\%) & \cellcolor{orange!25}3826 (13.8\%) \\
\bottomrule
\end{tabular}
% }
\end{table*}
\section{Experiments}
\label{sec:4_experiments}
\subsection{Evaluation Protocol and Baselines}
\label{sec:experiments_metrics}
\paragraph{\textbf{Metrics.}}
Since our tasks involve open-ended generation without unique ground-truth references, conventional N\mbox{-}gram-based metrics~\cite{bleu,rouge} are inadequate for faithfully evaluating output quality.  
We therefore adopt the JudgeLM paradigm~\cite{zhu2023judgelm}, in which a fine-tuned LLM serves as an automatic evaluator, assigning discrete quality scores (0--10) to model outputs across multiple dimensions, including relevance, factuality, and linguistic coherence.  
This evaluation method has been shown to closely align with human judgments, making it well-suited for assessing architectural VQA.
To tailor JudgeLM to our setting, we additional use a modified JudgeLM, by replacing the vanilla system prompt with an architecture-focused version that emphasizes grounding-encouraging the judge to prioritize architectural style, structural elements, and visual consistency. Full prompts and evaluation details are provided in the supplementary material.
In addition, we report standard text-matching metrics (GLEU~\cite{gleu}, METEOR~\cite{meteor}, ROUGE~\cite{rouge}); however, these serve only as auxiliary indicators. 
\paragraph{\textbf{Baselines.}}
We compare ArchGPT against representative open-source multimodal models of similar scale, including ShareGPT4V-7B~\cite{chen2024sharegpt4v}, Qwen-VL-Chat-7B~\cite{llm_5}, LLaVA-1.5-7B~\cite{li2024llava}, and InternVL3-8B~\cite{zhu2025internvl3}.  

\subsection{Quantitative Results}
\label{sec:exp_quantitative}

\begin{table}[b]
\centering
\caption{Comparisons with powerful baselines on several general-purpose LMM Benchmarks. 
With the first, second, and third values highlighted in red, orange, and yellow, respectively.
All the results of baselines are referred from~\cite{chen2024sharegpt4v}.}
\label{tab:robust_multimodal}
\begin{tabular}{lcccc}
\toprule
Model & MMB & LLaVA-W & MM-Vet & SQA \\
\midrule
InstructBLIP &36.0 & 60.9 &26.2& 60.5 \\
LLaVA-1.5 & \cellcolor{yellow!25}64.3 & \cellcolor{yellow!25}63.4 & \cellcolor{yellow!25}30.5 & 66.8 \\
Qwen-VL-Chat & 60.6 & - & - & \cellcolor{yellow!25}68.2 \\
ShareGPT4V & \cellcolor{orange!25}68.8 & \cellcolor{red!25}72.6 & \cellcolor{red!25}37.6 & \cellcolor{red!25}68.4 \\
ArchGPT (ours) & \cellcolor{red!25}69.0 & \cellcolor{orange!25}69.8 & \cellcolor{orange!25}35.7 & \cellcolor{orange!25}68.3 \\
\bottomrule
\end{tabular}
\end{table}
  % \cellcolor{red!25} \cellcolor{orange!25} \cellcolor{yellow!25}
% more detail InstructBLIP use Vicuna-7B model; Qwen-VL-Chat Qwen-7b language model. shareGPT4V, LLAVA-1.5 ARCHGPT vicuna-7B model
\begin{figure}[!htbp]
  \centering
\includegraphics[width=1.0\columnwidth]{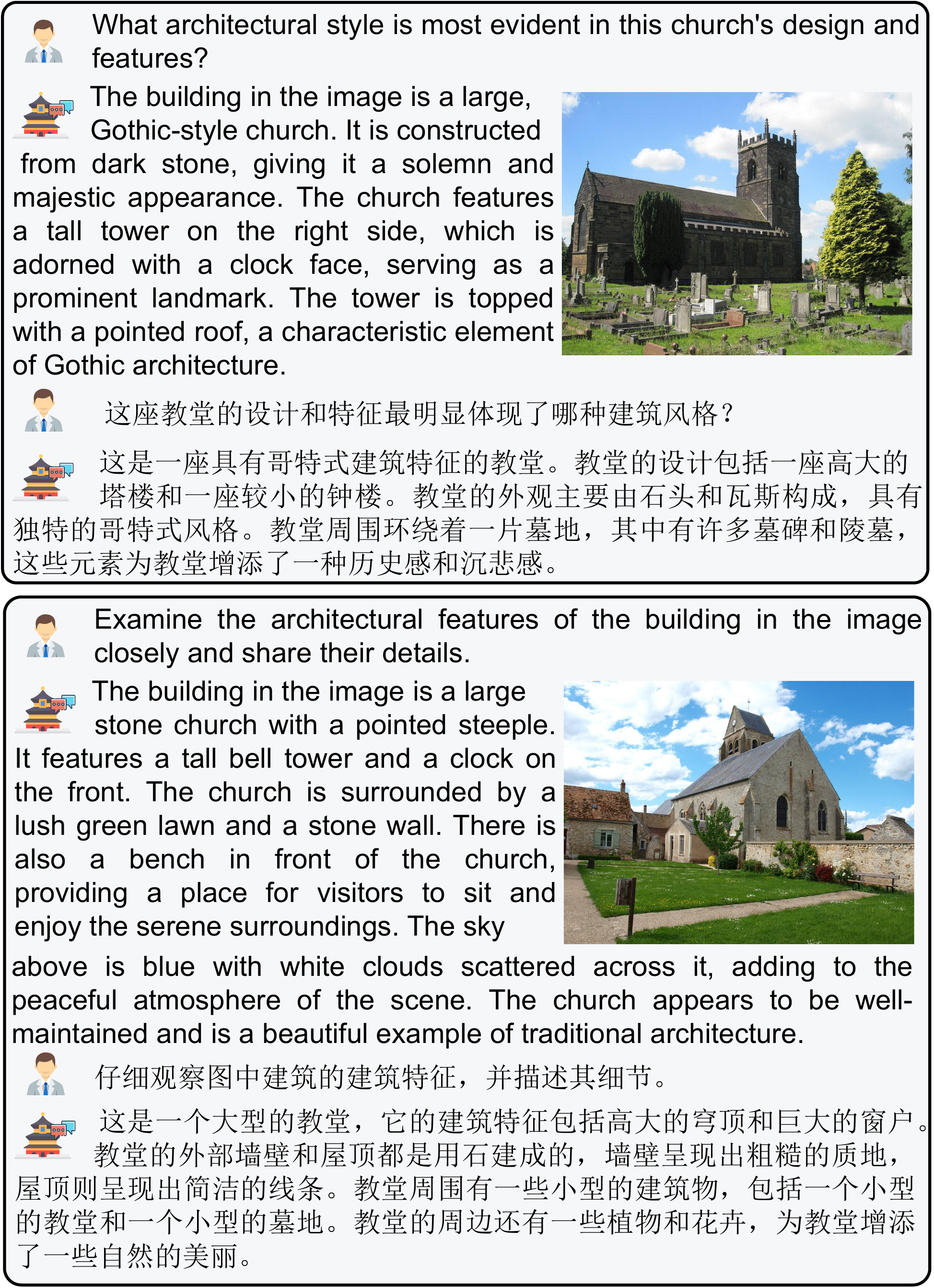}
\caption{Multilingual robustness analysis results. ArchGPT delivers consistent, high-quality responses to semantically equivalent questions in both English and Chinese.}
  \label{fig:exp_robust_multilingual}
\end{figure}
As shown in \cref{tab:quantitative_compare_main}, ArchGPT consistently surpasses strong open-source multimodal baselines across all evaluated metrics, demonstrating its superior capability in architecture-specific visual analysis.  
Notably, ArchGPT achieves a substantial +0.563-point improvement in the JudgeLM overall score $S_{judgeLM}$ over the next best model, InternVL3-8B. And +0.553-point improvement in the architectural specific JudgeLM overall score $S_{judgeLM}^*$ comapred to InternVL3-8B.

We further conduct a JudgeLM-based preference evaluation in which the judge compares anonymized answers from all methods to the same prompt and casts a single vote for the best response along four rubric dimensions: creativity, level of detail, logical consistency, and domain expertise. Candidate order is randomized and each item is evaluated under two permutations to mitigate position bias. As summarized in \cref{tab:judgelm_aspects}, ArchGPT attains the highest win rate across all five dimensions, indicating responses that are not only more informative and coherent but also exhibit stronger architectural expertise.% (Full voting prompt and rubric appear in the Supplementary Materials.)

\subsection{Qualitative Comparisons}

We present two cases for qualitative comparison: architectural-style recognition in \cref{fig:exp_qualitative_results_1} and detailed architectural analysis in \cref{fig:exp_qualitative_results_2}.  
In \cref{fig:exp_qualitative_results_1}, the monument is clearly an obelisk atop a stepped rectangular plinth with a stripped shaft-a classical commemorative form rendered with modern austerity.  
The baselines deviate in distinct ways: LLaVA~\cite{li2024llava} hallucinates a tetrastyle portico; ShareGPT-4V~\cite{chen2024sharegpt4v} invents flags and asserts unsupported material claims; QwenVL~\cite{llm_5} misclassifies it as Egyptian without citing corroborating motifs; and InternVL~\cite{zhu2025internvl3} offers generic observations lacking part-level grounding.
As shown in \cref{fig:exp_qualitative_results_2}, ArchGPT anchors its description in identifiable features-most notably the louvered belfry openings and a prominent side buttress-whereas other models remain superficial or introduce hallucinated elements (e.g., LLaVA: ribbed vaulting, high vantage, “obscured” spire; QwenVL: additional doors and windows). InternVL again produces vague and unsupported responses.
Overall, ArchGPT generates concise, verifiable descriptions grounded in observable cues, while the baselines tend to misclassify, hallucinate, or overlook architectural specificity.  
Additional qualitative comparisons are provided in the supplementary materials.

\subsection{Robustness and Generalization Analysis}

\paragraph{\textbf{General-purpose Multimodal Benchmarks.}}
To assess out-of-domain robustness after domain adaptation, we evaluate ArchGPT in zero-shot mode on widely used multimodal benchmarks, including MMB~\cite{mmbench}, LLaVA-W~\cite{llava_benchmark}, MM-Vet~\cite{mmvet}, and ScienceQA (SQA)~\cite{sqa}, following their official evaluation protocols and identical decoding settings to the baselines.
As summarized in \cref{tab:robust_multimodal}, ArchGPT shows only minor fluctuations relative to ShareGPT4V-7B and other 7B-8B models-sometimes slightly higher, sometimes slightly lower-while maintaining comparable multimodal understanding.
This indicates that our specialization does not induce catastrophic forgetting.
We attribute the stability to training lightweight LoRA adapters while freezing the backbone and keeping instruction formats aligned with the base model, which preserves general capabilities while adding strong architectural competence.
\begin{figure*}[!htbp]
  \centering
\includegraphics[width=1.86\columnwidth]{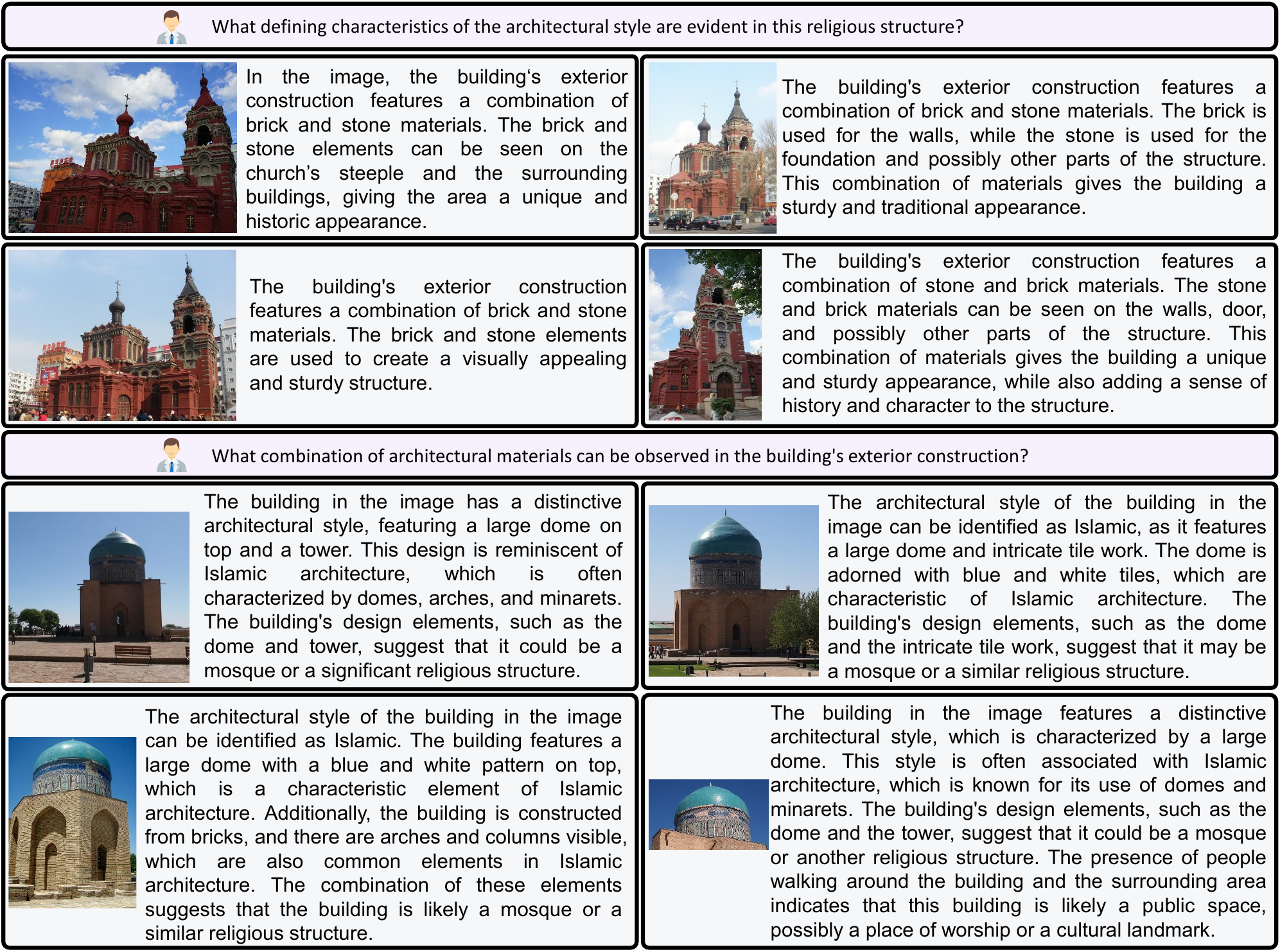}
\caption{Multiple views consistency analysis results. ArchGPT consistently identifies key architectural features and provides coherent descriptions across diverse images under varying viewpoints, lighting conditions, and times of day.}
  \label{fig:exp_robust_multiview}
\end{figure*}

\paragraph{\textbf{Multilingual Consistency Analysis.}}

We further assess the multilingual capabilities of ArchGPT, even though no multilingual VQA data were used during supervised fine-tuning.  
As shown in \cref{fig:exp_robust_multilingual}, we query ArchGPT with semantically equivalent questions in both English and Chinese.  
Importantly, ArchGPT delivers consistently high-quality responses across languages.  
We demonstrate more multilingual results in the supplementary materials.

\begin{table}[!htbp]
\centering
\caption{Ablation Studies. "CF", "FF", "LKC", and "WDE" denote "Coarse Filtering", "Fine Filtering", "LLM-Known Checker", and "WIKI-Data Extractor", respectively. $S_{judgeLM}$ and $S_{judgeLM}^{*}$ represent the JudgeLM scores without and with architectural emphasis in the system prompt.}
\label{tab:ablations}
\begin{tabular}{cccccc}
\toprule
w. CF & w. FF & w. LKC & WDE & $S_{judgeLM}$ & $S_{judgeLM}^{*}$ \\
\midrule
$\times$ & $\times$ & $\times$ & $\times$ & 6.079 &5.234  \\
$\times$ & $\times$ & $\checkmark$ & $\times$ & 6.171 &5.497  \\
$\times$ & $\times$ & $\times$ & $\checkmark$ & 6.661 &5.735  \\
$\times$ & $\times$ & $\checkmark$ & $\checkmark$ & 6.673 &5.818  \\
$\checkmark$ & $\times$ & $\checkmark$ & $\checkmark$ &7.394  & 6.723 \\
$\checkmark$ & $\checkmark$ & $\checkmark$ & $\checkmark$ &\textbf{7.713}  & \textbf{7.107} \\
\bottomrule
\end{tabular}
\end{table}

\paragraph{\textbf{Multiple Views Consistency Analysis.}}

To illustrate ArchGPT's potential for consistent reasoning across different perspectives, we design a multi-view consistency experiment.  
For each architectural scene, we consider sets of unconstrained images captured under varying viewpoints, lighting conditions, and times of day.  
We then pose identical questions across all images of the same scene, examining whether ArchGPT can provide semantically coherent responses despite visual variability.  
This setup offers a practical demonstration of the model's robustness in multi-view architectural understanding.
As shown in \cref{fig:exp_robust_multiview}, ArchGPT consistently identifies key architectural features and provides coherent descriptions across diverse images of the same structure.
We demonstrate more multiple views results in the supplementary materials.
\begin{figure}[!htbp]
  \centering
\includegraphics[width=1.0\columnwidth]{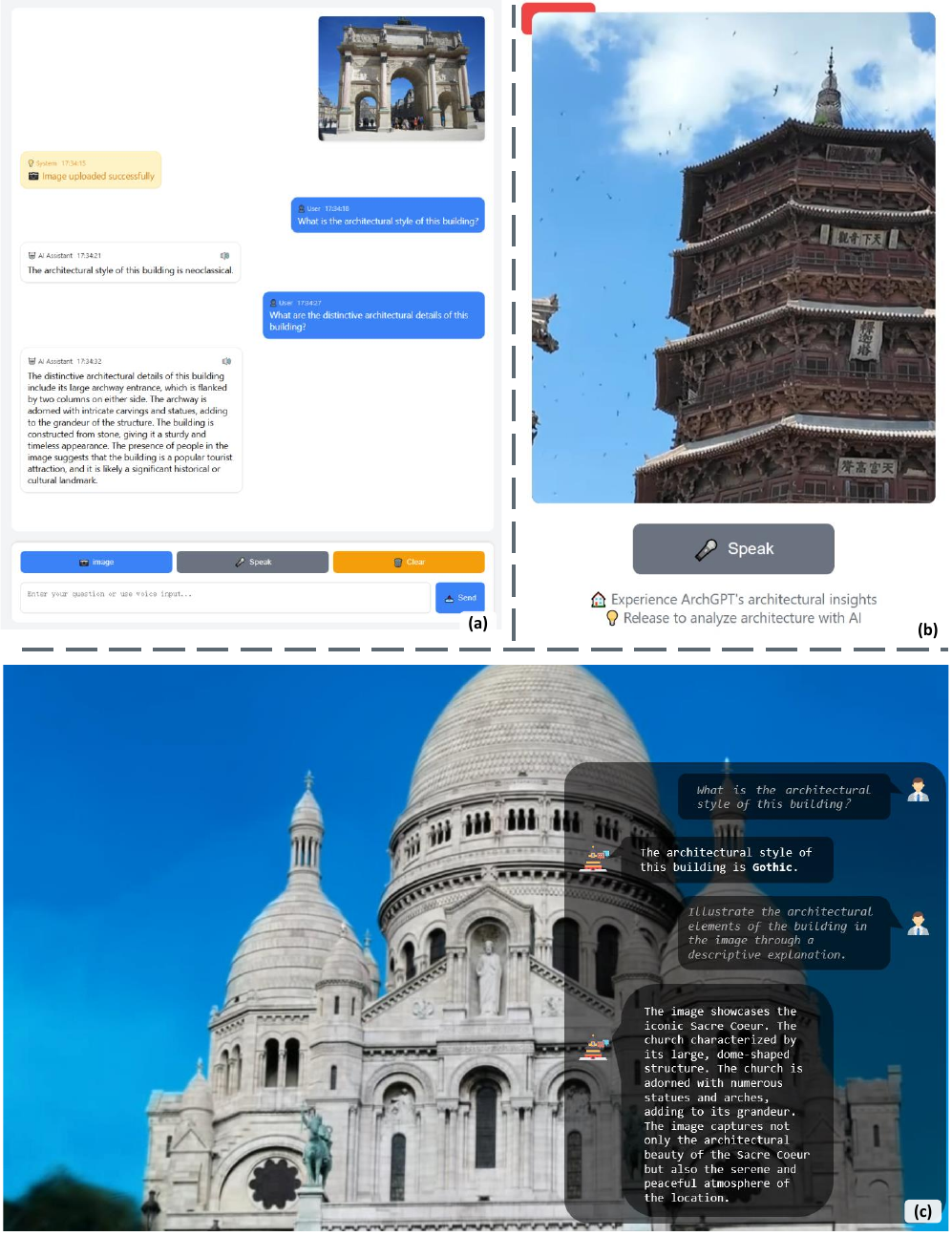}
\caption{Applications. The proposed ArchGPT can be seamlessly integrated into XR applications, such as (a) Interactive architectural conversational assistant, (b) AR-enabled immersive architectural interpretation, and (c) ask-as-your-explore virtual walkthroughs.}
  \label{fig:exp_app}
\end{figure}
\subsection{Ablation Studies}

We perform various ablation studies to evaluate the impact of different dataset-construction choices on ArchGPT's performance. All experiments use the same test split and maintain identical training samples, epochs, model parameters, and optimization settings to ensure a fair comparison. As shown in \cref{tab:ablations}, omitting the LLM-known checker (line 2 vs. line 1) noticeably degrades performance, as unchecked scenes introduce hallucinated or inconsistent annotations. Incorporating the Wiki-data extractor (line 3) improves textual supervision quality, resulting in higher scores. Finally, Lines 4-6 compare image-selection strategies: random sampling, coarse filtering, and the combined coarse-to-fine filtering. The results confirm that our coarse-to-fine approach effectively selects high-quality, occlusion-free images, yielding the best performance among the evaluated strategies.

\subsection{Applications}
We present three representative interactive applications of architectural analysis powered by the proposed ArchGPT, demonstrating its seamless integration into VR/MR/AR scenarios.

\paragraph{\textbf{Interactive Architectural Conversational Assistant.}}  

As shown in \cref{fig:exp_app} (a), we developed a web-based architectural conversational assistant. ArchGPT functions as a real-time tool that allows users to query architectural elements and receive context-aware explanations. This application highlights the potential of ArchGPT for interactive learning of architectural knowledge. 

\paragraph{\textbf{AR-Enabled Architectural Interpretation.}}  
As shown in \cref{fig:exp_app} (b), we developed an AR system that captures visual input in real time, with ArchGPT enabling users to pose interactive queries and experience AR-enabled, immersive architectural interpretation.

\paragraph{\textbf{Ask-as-Your-Explore Virtual Walkthroughs.}}  

Recent methods such as WildGaussians~\cite{kulhanek2024wildgaussians} and Look-at-the-Sky~\cite{wang2025look} enable high-quality, immersive free-viewpoint navigation of architectural scenes. ArchGPT is orthogonal to these approaches and can be seamlessly integrated into them. We demonstrate this by combining ArchGPT with Look-at-the-Sky: as shown in \cref{fig:exp_app} (c), during virtual walkthroughs, users can engage in interactive question answering from specific viewpoints, opening new possibilities for online exploration and learning of architectural knowledge, and paving the way for more engaging and scalable virtual experiences in the future.

We provide additional video demonstrations of these applications in the supplementary materials. We encourage readers to review the video results provided in the supplmentary materials.

\section{Limitations}
\label{sec:5_limitations}
While ArchGPT demonstrates strong performance, several limitations remain.  
First, the construction of the Arch-300K dataset relies on LLM knowledge priors; for remote or sparsely documented buildings, the lack of textual metadata constrains the LLM's ability to generate detailed and accurate descriptions.  
Second, although Arch-300K is sufficient for fine-tuning 7B LMMs, it is relatively small for training larger commercial models.  
These limitations could be mitigated in the future, as the proposed dataset construction paradigm is inherently scalable: by generating additional high-quality annotations, it may become possible to train larger multimodal models which-following established scaling laws-could achieve more specialized architectural VQA capabilities.

\section{Conclusion}
\label{sec:5_conclusion}
% In this paper, we introduced ArchGPT, a domain-adapted multimodal model for architecture-specific visual question answering, alongside Arch-300K, a large-scale dataset with fine-grained annotations covering stylistic, structural, symbolic, and contextual aspects of architectural imagery. To address the scarcity of domain-specific data and mitigate LLM-biased visual hallucination, we designed an LLM-assisted data construction pipeline that leverages web metadata, coarse-to-fine image filtering, and structured analysis generation. Extensive quantitative and qualitative evaluations demonstrate that ArchGPT effectively performs detailed architectural reasoning, outperforms strong multimodal baselines, and maintains robustness across diverse scenarios. Moreover, we showed how ArchGPT can be seamlessly integrated into AR and VR applications, enabling interactive exploration, guided interpretation, and cultural engagement. We believe our dataset, model, and pipeline provide a valuable foundation for scalable, architecture-aware XR experiences, and pave the way for future research in domain-specialized multimodal intelligence.
In this paper, we introduced ArchGPT, a domain-adapted multimodal model for architecture-specific visual question answering, together with Arch-300K, a large-scale dataset containing fine-grained annotations of stylistic, structural, symbolic, and contextual aspects of architectural imagery. Specifically, we designed an LLM-assisted data construction pipeline that combines web metadata, coarse-to-fine image filtering, and structured analysis generation. Extensive quantitative and qualitative evaluations show that ArchGPT achieves detailed architectural reasoning, surpasses strong multimodal baselines, and remains robust across diverse scenarios. Furthermore, we demonstrated that ArchGPT can be seamlessly integrated into AR and VR applications, enabling interactive conversation, AR-enabled interpretation, and ask-as-your-explore virtual walkthroughs. Overall, our dataset, model, and pipeline establish a foundation for scalable, architecture-aware VR experiences and pave the way for future research in domain-specialized multimodal intelligence.

\section*{Supplemental Material}
\label{sec:supplemental_materials}
\setcounter{figure}{0}
\setcounter{table}{0}
\renewcommand\thesection{\Alph{section}}
\renewcommand\thetable{\Alph{table}}
\renewcommand\thefigure{\Alph{figure}}

In this supplementary material, we provide an in-depth explanation of our proposed Arch-300K dataset in Sec. A, more about the dataset implementation details in Sec. B, more  about the large multimodal model implementation detail in Sec. C, and demonstrating more experiment details and results in Sec. D.
\section{More Details about the Arch-300K Dataset}
\label{sec:appendix-arch300k}
\begin{figure}[!htb]
\centering
\includegraphics[width=1.0\columnwidth]{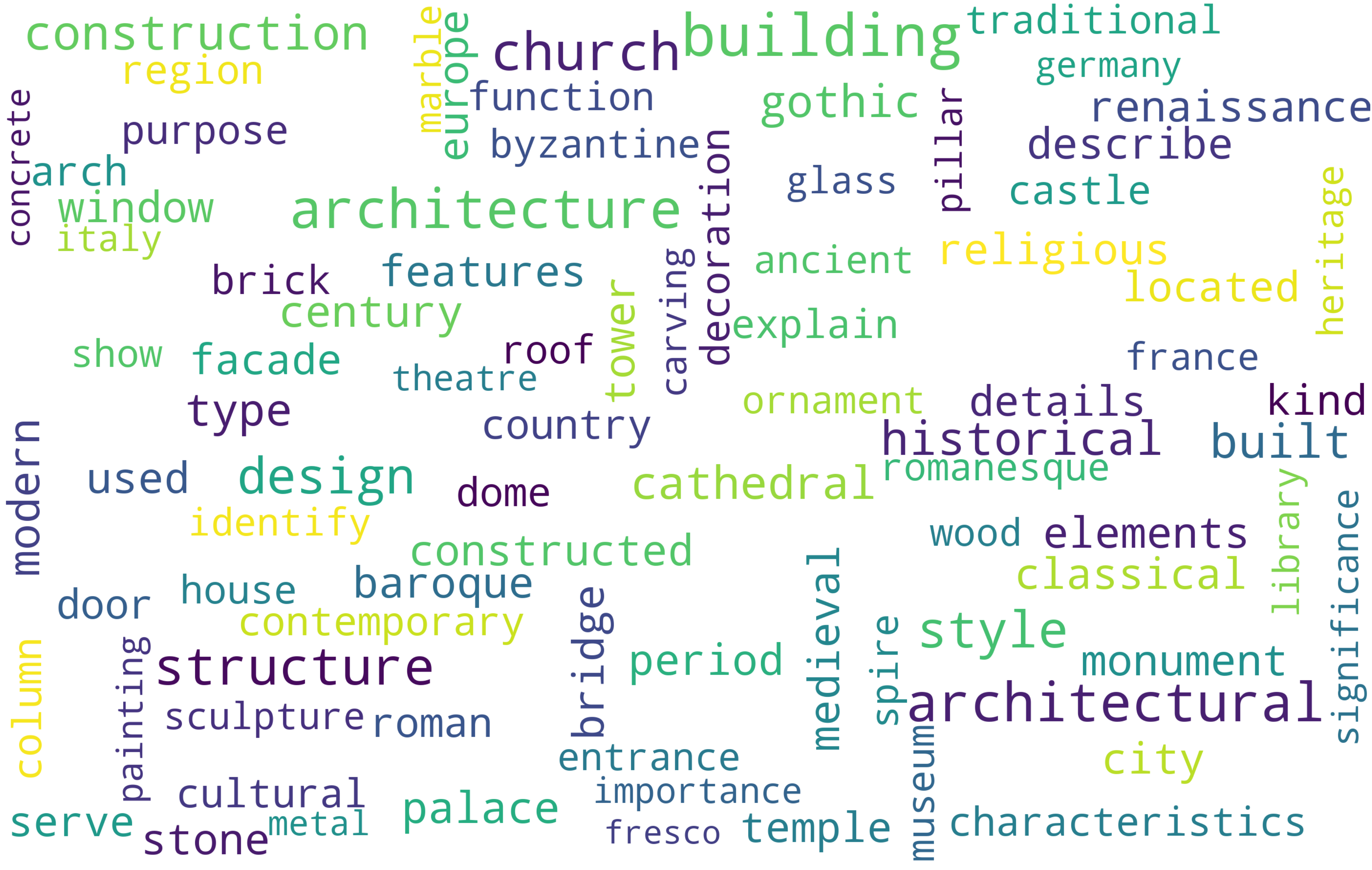}
\caption{Question word cloud for the Arch-300K dataset.}
\label{fig:fig_sup_wordcloud_question}
\end{figure}

\begin{figure}[!htb]
\centering
\includegraphics[width=1.0\columnwidth]{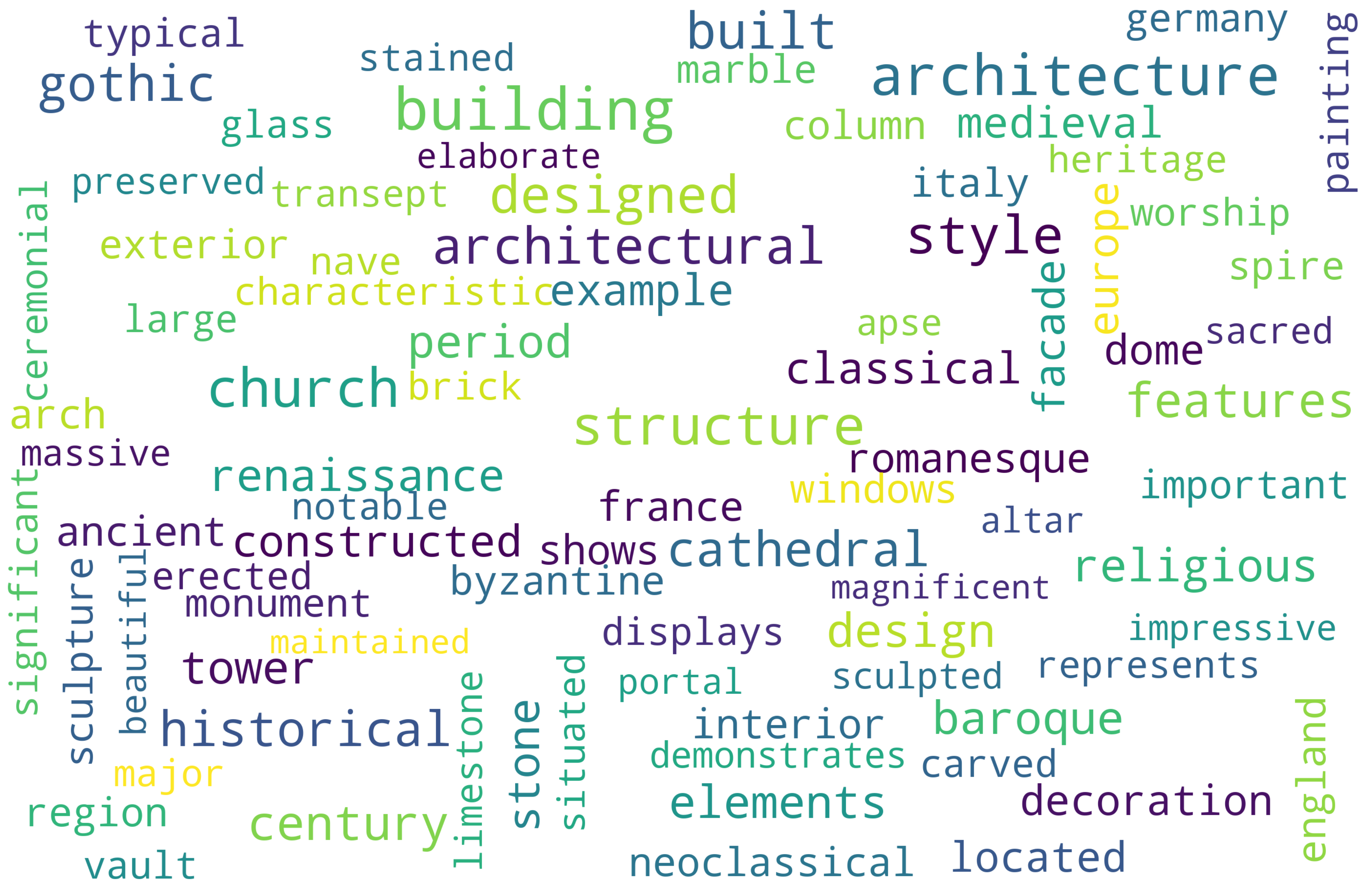}
\caption{Answer word cloud for the Arch-300K dataset.}
\label{fig:fig_sup_wordcloud_answer}
\end{figure}

\begin{table}[h]
  \centering
  \caption{Arch-300K dataset summary statistics.}
  \label{tab:arch300k-stats}
  \begin{tabular}{l r}
    \hline
    Total VQA items & 315{,}247 \\
    Unique scenes & 8{,}643 \\
    Max images per scene & 8 \\
    Avg.\ questions per scene & 72.9 \\
    Question type: detailed description & 23\% \\
    Question type: aspect-guided conversation & 77\% \\
    Avg.\ answer length (words) & 81.7 \\
    \hline
  \end{tabular}
\end{table}
\cref{tab:arch300k-stats} summarizes key statistics of the Arch-300K dataset. Arch-300K comprises 315{,}247 VQA items across 8{,}643 unique architectural scenes, averaging 72.9 questions per scene, with the number of images per scene capped at 8. The corpus spans a broad range of architecture types-including, but not limited to, church, monument, palace, lighthouse, rathaus (town hall), mosque, temple, and synagogue-covering varied periods, styles, functions, and cultural contexts. The question set mixes two complementary query modes: roughly 23\% target detailed description generation, encouraging paragraph-level, holistic characterization of a facade or structure, while the remaining 77\% are aspect-guided conversation prompts that focus on specific facets (e.g., stylistic cues, structural elements, ornamentation, materials, context). \cref{fig:fig_sup_wordcloud_question,fig:fig_sup_wordcloud_answer} visualize the lexical distribution as word clouds, where font size is proportional to token frequency (larger words occur more often) and common stopwords are removed. Answers are intentionally long-form to encourage evidence-backed reasoning and verifiable claims; the average answer length is 81.7 words, favoring grounded explanations beyond terse labels.

\section{More Implementation Details about the Arch-300K dataset.}
\subsection{Wikimedia Commons and MegaScenes Preprocessing}
As discussed in the main paper, we treat Wikimedia Commons~\cite{wikimediacommons} categories as distinct scenes, following the MegaScenes~\cite{megascenes} protocol. Concretely, a scene corresponds to a unique architectural entity (a single building or complex), and synonymous or alias categories referring to the same entity are merged. Wikimedia Commons is a free, collaborative media repository that hosts images, video, and audio. Beyond human-readable descriptions, files carry machine-readable Structured Data on Commons (SDC) built on Wikibase~\cite{vrandevcic2014wikidata}, enabling multilingual, queryable metadata (e.g., depicts, creator, license) and organization via a large category taxonomy. Commons categories and galleries are linked to corresponding Wikidata items through sitelinks, providing consistent identifiers across languages.
MegaScenes initializes candidate scenes by collecting Wikidata entries that link to Commons categories and retain those whose Commons-Wikidata links are cyclic (the Commons category links back to the same Wikidata item), which removes overly broad or thematic categories (e.g., “Fountains”). It further discards entries that are exclusively GLAM instances (galleries, libraries, archives, museums) to avoid 2D scans and other non-photographic media.
For image acquisition, it starts from each accepted Commons category and recurses into subcategories to a maximum depth of four. To keep the crawl on-topic, it excludes subcategories that match a curated blacklist of diverging patterns (e.g., "People associated with …"), and it requires the subcategory name to contain the original Commons category name, the linked Wikidata label, or any recorded alias.

In our work, we do not re-implement the crawl; instead, we directly reuse the scene list and downloaded image sets produced by MegaScenes, and then apply our architecture-focused filtering and VQA construction on top of these curated categories and images.

\subsection{More Details on the Coarse-to-Fine Image Filtering}

We adopt a two-stage pipeline. In the coarse stage, we apply VGGT~\cite{vggt} to each scene to derive geometry-driven cues and coarse region masks used for downstream filtering. To avoid GPU out-of-memory (OOM) issues in scenes with very large photo collections, we cap the number of input images per scene at 200; for scenes exceeding this limit, we uniformly sample 200 images at random and feed them to VGGT. This simple cap keeps memory bounded while retaining sufficient multi-view coverage for reliable coarse filtering.

In the fine stage, we use the standard Segment Anything Model \cite{sam} (SAM, ViT-H backbone) as a promptable segmenter. Positive point prompts are uniformly sampled within the coarse masks produced by the previous stage, while negative prompts were empirically found to provide little benefit for mask precision in our setting. We therefore omit negative prompts and set the number of point prompts to 10, which we observed to work best across scenes. All other components follow their default configurations unless otherwise specified.

\subsection{More details about the LLM-Guided Text Verification and knowledge distillation.}
As discussed in the main paper, we employ Gemini 2.5 Pro~\cite{team2023gemini} for LLM-guided text verification and knowledge distillation. Below we present the exact prompts used in our pipeline. For textural metadata analysis, the full prompt is:

\begin{quote}\small\ttfamily
Please analyze the following wikidata information about the \textcolor{red}{\{scene\_name\}}, and return a concise description  that focuses on the architecture's key characteristics and architectural features. Please DO NOT incorporate your own knowledge; the extraction should be ONLY based entirely on the material provided. If you don't know, just response 'Unknown'. The material is as follows: \textcolor{red}{\{wiki\_raw\_data\}}
\end{quote}

For extracting the refined architecture's name, the prompt is:

\begin{quote}\small\ttfamily
This architecture's raw name is: \textcolor{red}{\{scene\_name\}}. Please refine the architecture's name from the following wikidata. Please DO NOT incorporate your own knowledge; the extraction should be ONLY based entirely on the material provided. If you don't know, just response 'Unknown'. The material is as follows: \textcolor{red}{\{wiki\_raw\_data\}}"
\end{quote}
We empirically observe that extracting a architecture's construction time and location exclusively from raw Wikidata marginally reduces LLM hallucinations and improves accuracy. For extracting country and year, the full prompt is:

\begin{quote}\small\ttfamily
Please analyze the following wikidata and return the architecture's country (such as China) and year (such as 1999). Please DO NOT incorporate your own knowledge; the extraction should be ONLY based entirely on the material provided. The material is as follows: Architecture name: \textcolor{red}{\{scene\_formal\_name\}}. Introduction: \textcolor{red}{\{wiki\_raw\_data\}}. The answer format should be: Country, Year (Such as: China, Unknown)"
\end{quote}

For detailed-description question design, we take inspiration from LLaVA~\cite{llava_benchmark} to construct a diverse set of question templates and randomly select one at runtime. The question list is:

\begin{quote}\small\ttfamily
\quad "Describe the architectural features and design elements of the architecture in the following image in detail.",\\
\quad "Provide a detailed description of the architectural characteristics of the building shown in the given image.",\\
\quad "Explain the various details of the architectural design you can observe in the image.",\\
\quad "Share a comprehensive analysis of the architectural features presented in the image.",\\
\quad "Offer a thorough analysis of the building's architectural elements visible in the image.",\\
\quad "Explain the various architectural details and design aspects of the building in the displayed image with great detail.",\\
\quad "Characterize the architectural features of the building in the image using a well-detailed description.",\\
\quad "Break down the architectural elements of the architecture in the image in a detailed manner.",\\
\quad "Walk through the important details of the architectural design visible in the image.",\\
\quad "Portray the architectural features and design elements of the architecture in the image with a rich, descriptive narrative.",\\
\quad "Narrate the architectural characteristics of the architecture in the image with precision.",\\
\quad "Analyze the architectural design and features of the architecture in the image in a comprehensive and detailed manner.",\\
\quad "Illustrate the architectural elements of the architecture in the image through a descriptive explanation.",\\
\quad "Examine the architectural features of the building in the image closely and share their details.",\\
\quad "Write an exhaustive depiction of the architectural characteristics of the architecture in the given image.",\\
\quad "Carefully observe the building in the image and share the details of its architectural design and features."\\
\end{quote}

Given a selected question, the matching-answer prompt is:

\begin{quote}\small\ttfamily
You are a professional architect. Please answer the following question about the \textcolor{red}{\{scene\_formal\_name\}}: \textcolor{red}{\{detailed\_description\_question\}}. Please compose a coherent paragraph, approximately \textcolor{red}{\{detailed\_words\}} words. Start your response with 'In this image, I can see...' or 'This image shows...' Only focus on the VISUAL CHARACTERISTICS. DO NOT mention the specific architecture name. You may use conjunctions as needed to ensure coherence in semantics and logic in your analysis. Just response my question, do not explain your reason. Supplementary material: The architecture's name:\textcolor{red}{\{scene\_formal\_name\}}, location: \textcolor{red}{\{scene\_location\}},construct year:\textcolor{red}{\{scene\_year\}}, other description:\textcolor{red}{\{scene\_refined\_description\}}
\end{quote}

Accordingly, we also use the following prompt to generate a matching answer:

\begin{quote}\small\ttfamily
You are a professional architect. Please answer the following question about the \textcolor{red}{\{scene\_formal\_name\}}: \textcolor{red}{\{detailed\_description\_question\}}. Please compose a coherent paragraph, approximately \textcolor{red}{\{detailed\_words\}} words. Start your response with 'In this image, I can see...' or 'This image shows...'. Only focus on the VISUAL CHARACTERISTICS. DO NOT mention the specific architecture name. You may use conjunctions as needed to ensure coherence in semantics and logic in your analysis. Just response my question, do not explain your reason."
\end{quote}

For aspect-guided conversation, the aspect-selection prompt is:

\begin{quote}\small\ttfamily
You are a professional architect. Please select up to five essential aspects for the formal analysis of the \textcolor{red}{\{scene\_formal\_name\}}. You may choose fewer if preferred. The list is: [architectural style, architectural elements, architectural details, architectural context, architectural innovation, architectural symbolism, architectural materials]. You also can propose other aspects. Please list the aspects in descending order of importance in a List. Just response the aspect name, dont explain your reason. Your output should be in the format'[YOUR SELECTED CHARACTERISTICS]', such as '[architectural style, architectural elements]' Supplementary material: The architecture's name:\textcolor{red}{\{scene\_formal\_name\}}, location: \textcolor{red}{\{scene\_location\}},construct year:\textcolor{red}{\{scene\_year\}}, other description:\textcolor{red}{\{scene\_refined\_description\}}
\end{quote}

After parsing the selected aspects, the question-generation prompt for each aspect is:

\begin{quote}\small\ttfamily
You are a professional architect. Create a concise, clear question about the \textcolor{red}{\{aspect\}} of the architecture.  The question should be direct and professional, around 10-15 words. Start with phrases like 'What', 'How does', 'Describe', 'Which', etc. Example formats: 'What \textcolor{red}{\{aspect\}} can you observe in this architecture?', 'How does this architecture demonstrate \textcolor{red}{\{aspect\}}?' Do not mention the specific architecture name. Just provide the question directly. Supplementary material: The architecture's name:\textcolor{red}{\{scene\_formal\_name\}}, location: \textcolor{red}{\{scene\_location\}}, construct year:\textcolor{red}{\{scene\_year\}}, other description:\textcolor{red}{\{scene\_refined\_description\}}
\end{quote}

For generating the answers, the prompt is:

\begin{quote}\small\ttfamily
You are a professional architect. Please provide a detailed analysis of the \textcolor{red}{\{aspect\}} of the \textcolor{red}{scene\_formal\_name\}}. Start your response with 'In this image, the \textcolor{red}{\{aspect\}}' or 'This image shows that the \textcolor{red}{\{aspect\}}' Focus specifically on the \textcolor{red}{\{aspect\}} that can be visually observed in the image and provide approximately \textcolor{red}{\{aspect\_words\}} words. Use clear, professional language without markdown formatting. DO NOT mention the specific architecture name in your analysis. Just provide the analysis directly.
\end{quote}

\begin{figure}[!htb]
\centering
\includegraphics[width=1.0\columnwidth]{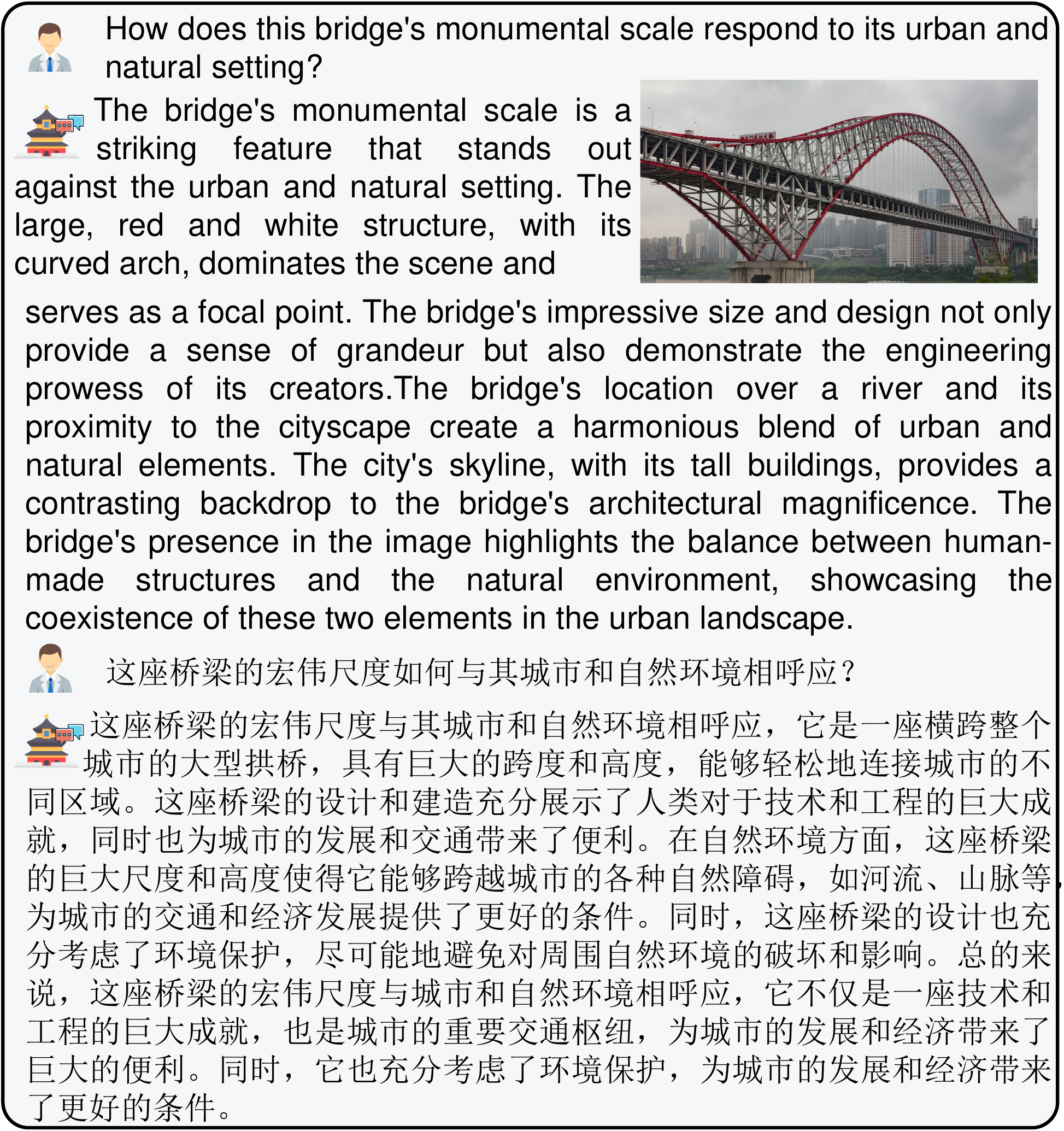}
\caption{Multilingual robustness analysis results.}
\label{fig:fig_sup_robust_language_2}
\end{figure}

\begin{figure}[!htb]
\centering
\includegraphics[width=1.0\columnwidth]{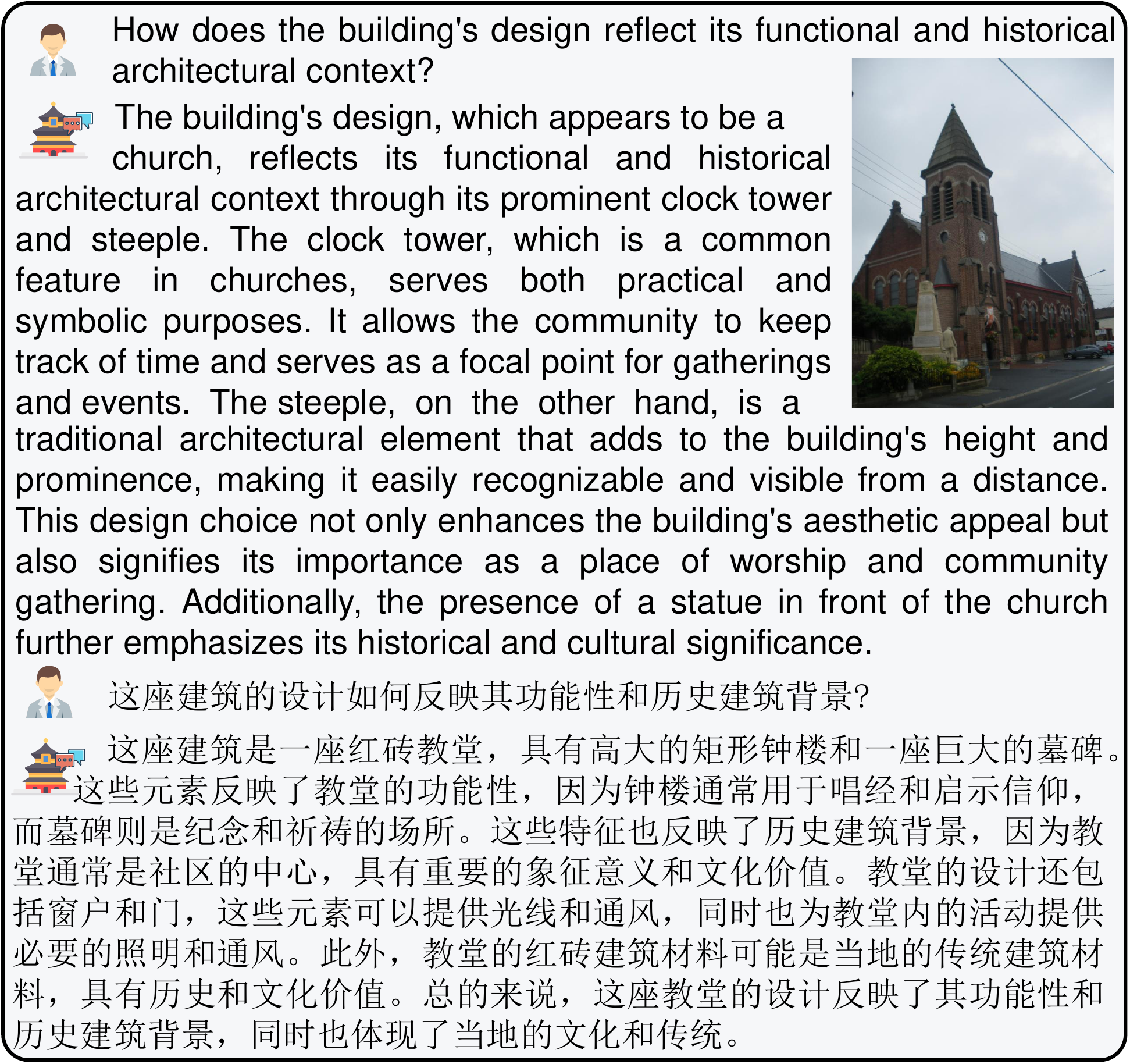}
\caption{Multilingual robustness analysis results.}
\label{fig:fig_sup_robust_language_3}
\end{figure}
\section{More Implementation Details of ArchGPT}
\label{app:model_details}
As shown in the main paper, we keep the backbone unchanged and record only the essentials for reproducibility: ArchGPT is trained by updating LoRA adapters on the LLM ($r{=}128$, $\alpha{=}256$) together with a two-layer MLP projector with GELU, while freezing both the base LLM and the vision tower. Supervision adopts a multi-annotation fusion scheme inspired by Molmo~\cite{deitke2025molmo}-per image, all descriptions and QA pairs are concatenated into a single sequence with appropriate attention masking to maximize context utilization. Optimization uses AdamW~\cite{adam} with a cosine schedule and $3\%$ warmup at a learning rate of $2{\times}10^{-5}$ for all trainable modules (weight decay $0.0$), with a maximum context length of $2048$ tokens. Training runs for two epochs in bfloat16 with TF32-enabled matmuls, gradient checkpointing, and DeepSpeed ZeRO-3~\cite{rasley2020deepspeed} on a single 40GB GPU.
\section{More Experiment Details and Qualitative Results.}

We report two JudgeLM configurations used in our quantitative comparisons: a vanilla setting and an architecture-tailored setting. Their system prompts are listed below.

\textbf{Vanilla JudgeLM system prompt:}
\begin{quote}\small\ttfamily
We would like to request your feedback on the performance of two AI assistants' answers in comparing with the reference answer and determine if they match meaningfully. Please rate the helpfulness, relevance, accuracy, level of details of the Assistants' answers. To accomplish the task, you must : 1. Focus on the meaningful match between the reference answer and the Assistants' answers. 2. Consider synonyms or paraphrases as valid matches. 3. Evaluate the correctness of the Assistants' answers compared to the reference answer. 4. If there are multiple reference answers, the Assistants' answer is considered correct as long as it is close to any of the answers.Please first output a single line containing only two values indicating the scores for Assistant 1 and 2, respectively. The two scores are separated by a space. In the subsequent line, please provide a comprehensive explanation of your evaluation, avoiding any potential bias and ensuring that the order in which the responses were presented does not affect your judgment.
\end{quote}

\textbf{Architecture-tailored JudgeLM system prompt:}
\begin{quote}\small\ttfamily
  You are an expert architect and architectural historian. We request your feedback on the performance of two AI assistants' answers by comparing them with the reference answer and determining whether they match meaningfully for architectural analysis. Please rate helpfulness, relevance, accuracy, and level of detail with emphasis on architectural style/typology, character-defining elements (arches, vaults, column orders, fenestration, roof/plan logic), materials and structural system, and site/context (period, region, function). To accomplish the task, you must: 1) focus on the architectural-semantic match between the reference and the assistants' answers; 2) treat accepted synonyms or paraphrases as valid (e.g., "Neo-Gothic" equals "Gothic Revival" when context fits); 3) evaluate correctness and penalize hallucinations, anachronisms, or misattributed elements; 4) if multiple reference answers are provided, an answer is correct if it aligns with any of them; 5) when the reference indicates insufficient information, reward cautious uncertainty over guessing. Please first output a single line containing only two integer scores (0-10) for Assistant 1 and Assistant 2, separated by a space. In the subsequent line, provide a concise, neutral explanation citing concrete architectural cues (e.g., pointed arches, ribbed vaults, ashlar masonry) and justifying any deductions, avoiding bias from response order.
\end{quote}

\begin{figure*}[!htb]
\centering
\includegraphics[width=2.0\columnwidth]{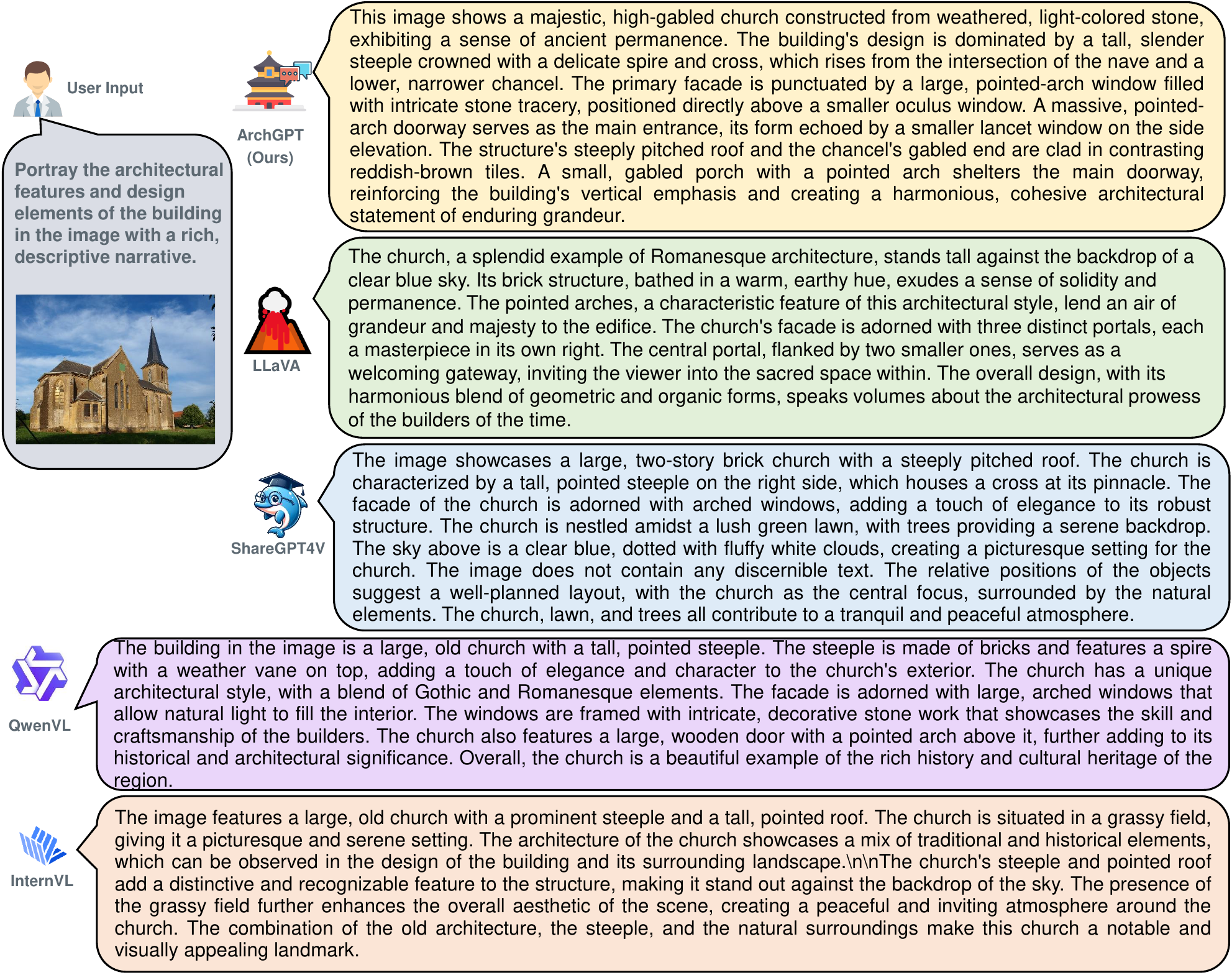}
\caption{Qualitative comparisons between ArchGPT and other LMMs.}
\label{fig:fig_sup_qualitative_results_3}
\end{figure*}

\begin{figure*}[!htb]
\centering
\includegraphics[width=2.0\columnwidth]{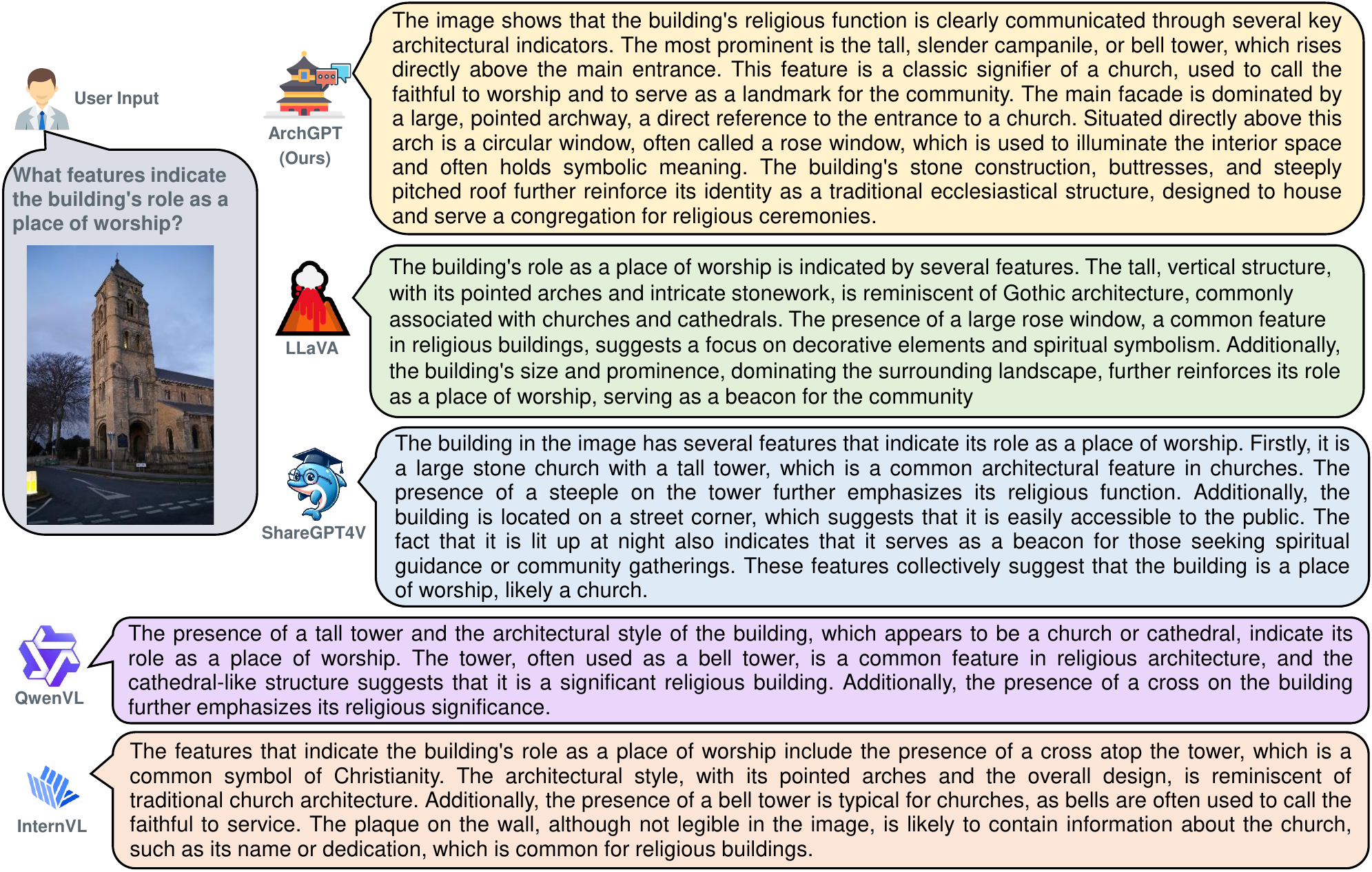}
\caption{Qualitative comparisons between ArchGPT and other LMMs.}
\label{fig:fig_sup_qualitative_results_6}
\end{figure*}

\begin{figure*}[!htb]
\centering
\includegraphics[width=2.0\columnwidth]{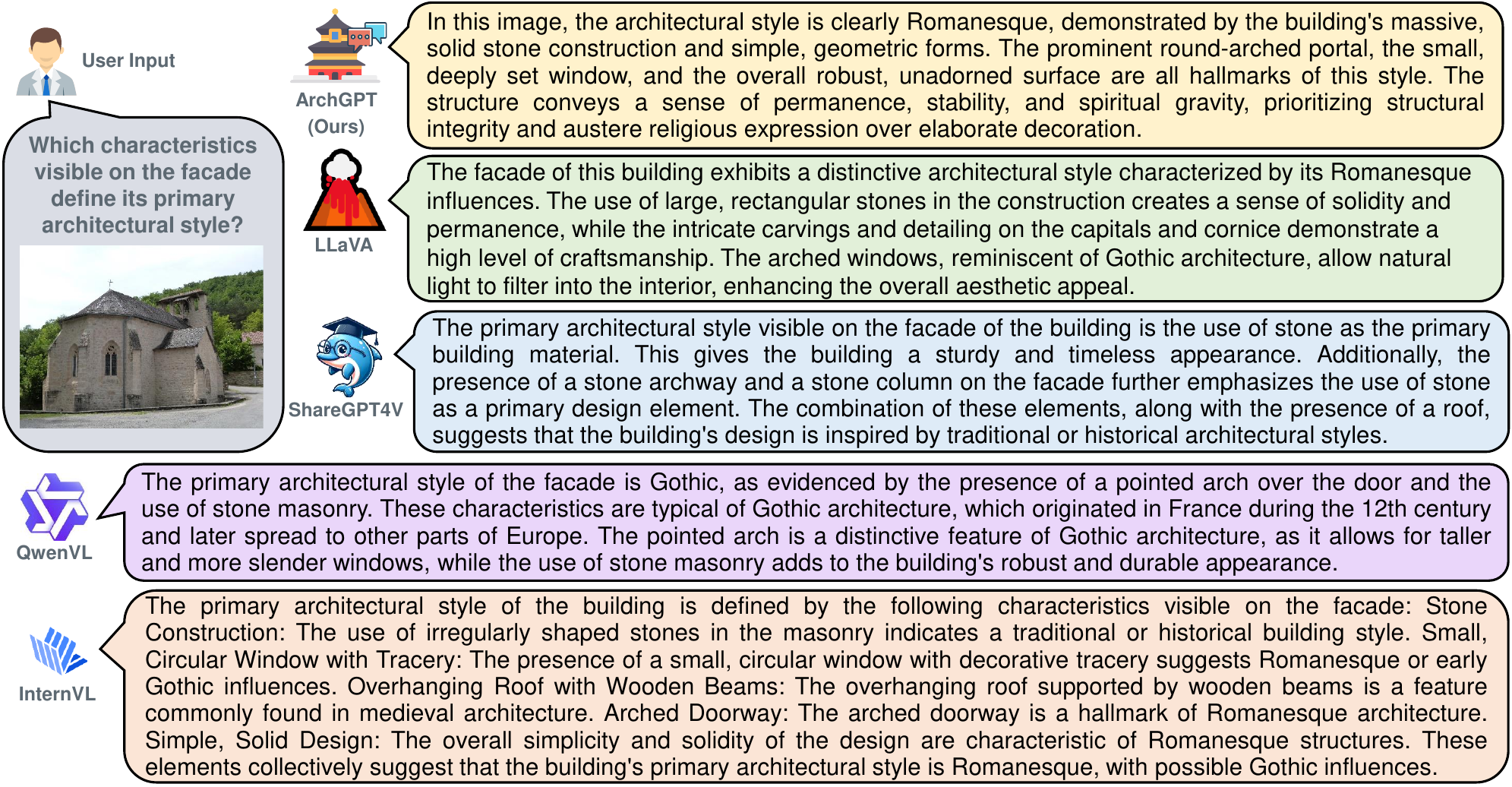}
\caption{Qualitative comparisons between ArchGPT and other LMMs.}
\label{fig:fig_sup_qualitative_results_4}
\end{figure*}

\begin{figure*}[!htb]
\centering
\includegraphics[width=2.0\columnwidth]{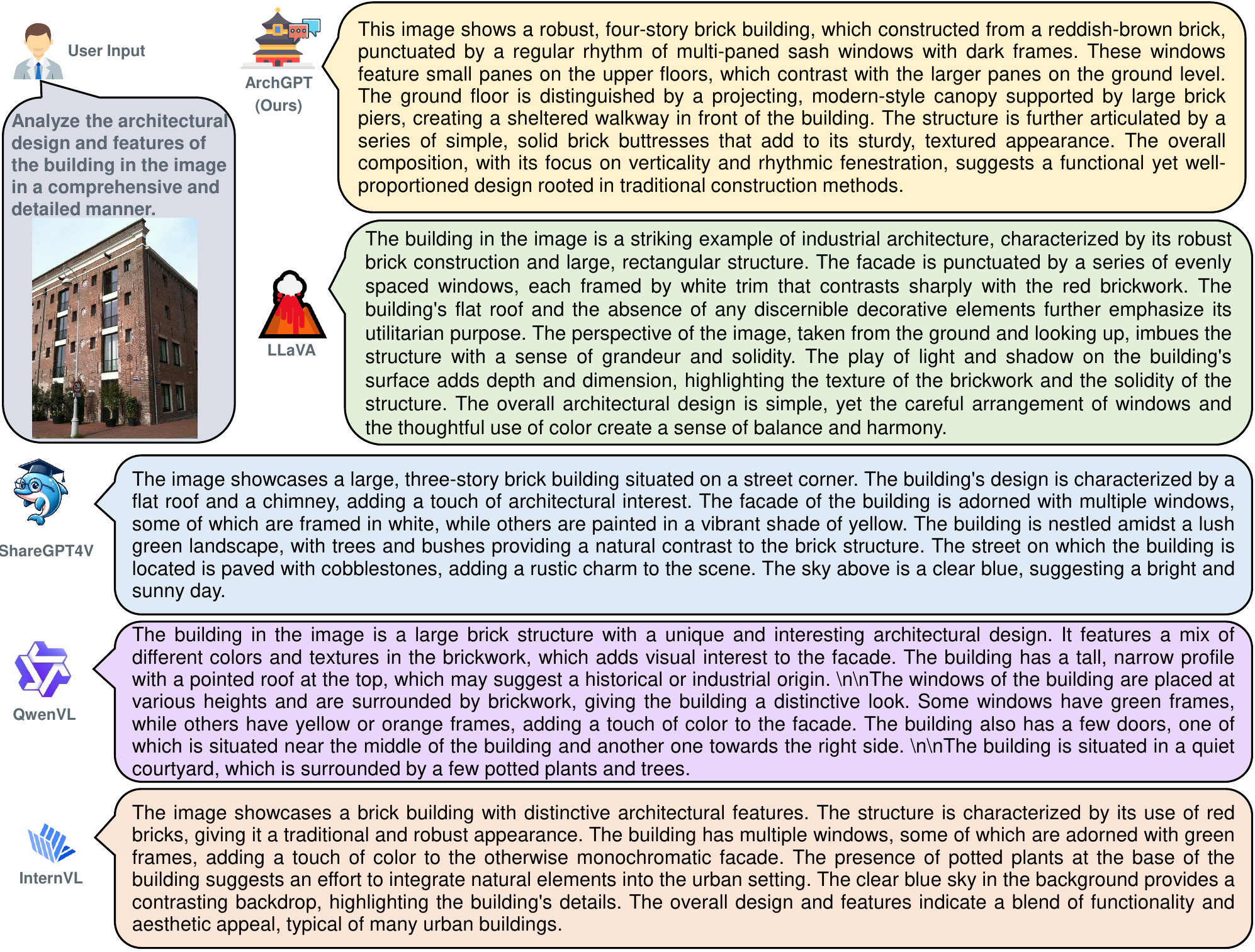}
\caption{Qualitative comparisons between ArchGPT and other LMMs.}
\label{fig:fig_sup_qualitative_results_5}
\end{figure*}

\begin{figure*}[!htb]
\centering
\includegraphics[width=2.0\columnwidth]{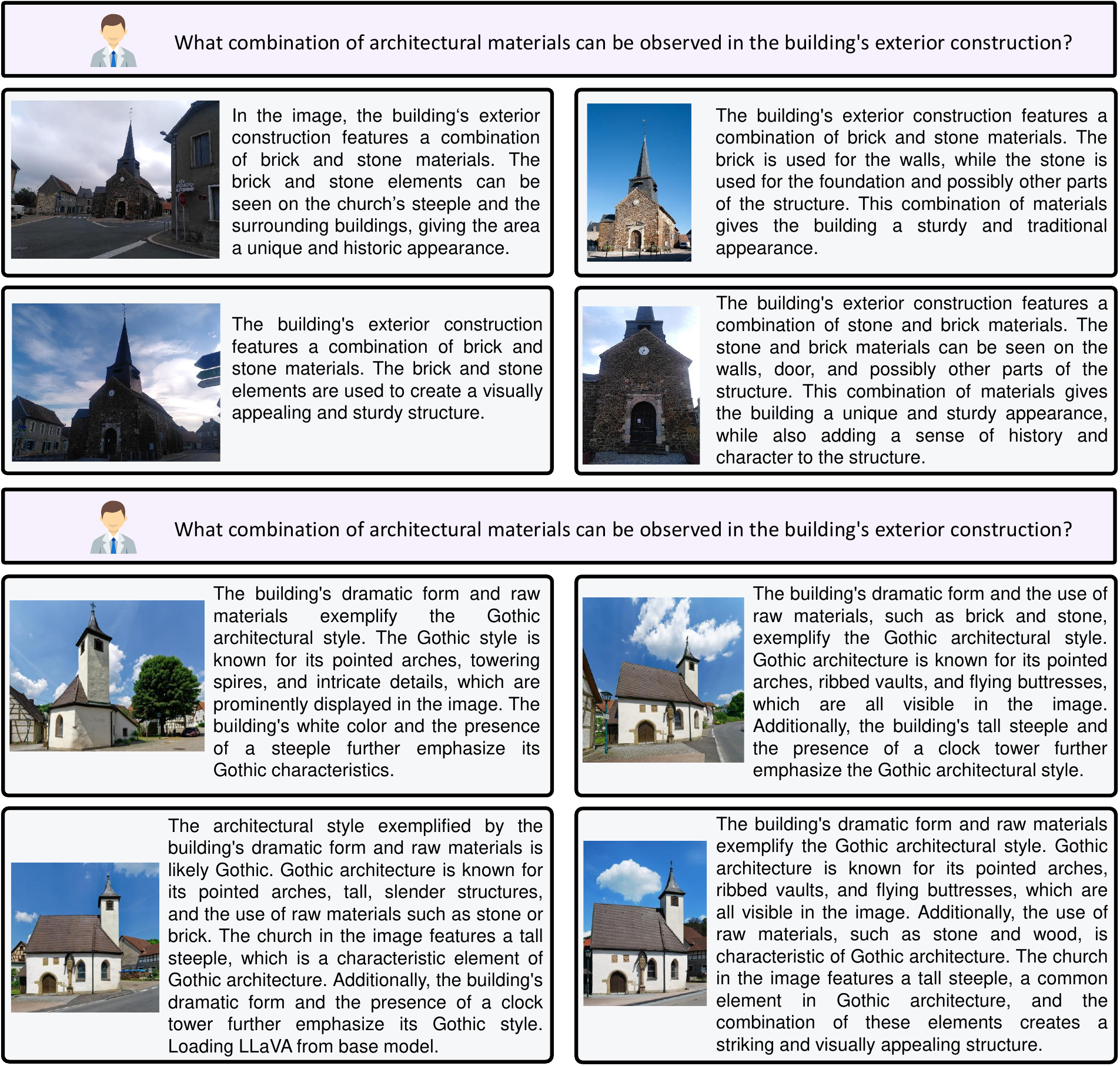}
\caption{Multiple view consistency analysis results. }
\label{fig:fig_sup_robust_mv_2}
\end{figure*}

Then, we present additional qualitative results for ArchGPT: \cref{fig:fig_sup_qualitative_results_3}, \cref{fig:fig_sup_qualitative_results_4}, \cref{fig:fig_sup_qualitative_results_5}, and \cref{fig:fig_sup_qualitative_results_6} show further qualitative comparisons, \cref{fig:fig_sup_robust_language_2} and \cref{fig:fig_sup_robust_language_3} report additional results from our multilingual robustness analysis. \cref{fig:fig_sup_robust_mv_2} presents additional results for the multi-view consistency analysis.

%\section*{Supplemental Materials}
%\label{sec:supplemental_materials}
%\section*{Figure Credits}
%\label{sec:figure_credits}
% \acknowledgments{
%   The authors wish to thank A, B, and C. This work was supported in part by
%   a grant from XYZ.}

%\bibliographystyle{abbrv}
\bibliographystyle{abbrv-doi}

\bibliography{2_reference}
\end{document}